\def\arxivcopy{1}
\documentclass[e-only,10pt,reqno]{ofj}

\usepackage{tensorNotation}
\usepackage{finiteVolumeNotation}

\usepackage{setspace}
\usepackage{lineno}
\usepackage{color}
\usepackage[dvipsnames,svgnames,x11names]{xcolor}
\usepackage[hang,flushmargin]{footmisc}
\usepackage{orcidlink}
\usepackage{cite}
\usepackage{xspace} % Don't remove spaces following \mbox{} etc

\ifdefined\review\else\ifdefined\arxivcopy\else
\usepackage{showlabels}

\fi\fi

\usepackage[colorlinks=true,linkcolor=black,anchorcolor=black,citecolor=black,filecolor=black,menucolor=black,runcolor=black,urlcolor=black]{hyperref}

\newcommand{\authorcontributions}[1]{%
\vspace{6pt}\noindent{\fontsize{9}{11.2}\selectfont\textbf{Author Contributions:} {#1}\par}}
\newcommand{\dataavailability}[1]{%
\vspace{6pt}\noindent{\fontsize{9}{11.2}\selectfont\textbf{Data Availability:} {#1}\par}}

\let\paragraph\undefined

\renewcommand{\datename}{$^{*}$ Corresponding author}
\copyrightinfo{2026}{The Authors. This work is an open access article under the {\color{blue}{\href{https://creativecommons.org/licenses/by-sa/4.0/}{CC BY-SA 4.0}}} license}
\newcommand{\OF}[0]{OpenFOAM\textsuperscript{\textregistered}\xspace}

\usepackage{amssymb} % \lesssim, \gtrsim
\usepackage{float}   % [H] float placement
\usepackage{placeins} % \FloatBarrier where needed

\IfFileExists{ulem.sty}%  (not in every TeX installation, e.g. the CI image)
  {\usepackage[normalem]{ulem}}
  {}

\newcommand{\coderepo}{figures}

\ifdefined\review
\DeclareRobustCommand{\datasource}[1]{\nolinkurl{#1}}
\else\ifdefined\arxivcopy
\DeclareRobustCommand{\datasource}[1]{\href{https://github.com/isoAdvector/fluxinator/tree/main/#1}{\nolinkurl{#1}}}
\else
\DeclareRobustCommand{\datasource}[1]{\href{run:../fluxinator/#1}{\nolinkurl{#1}}}
\fi\fi
\newcommand{\datanote}[1]{\textit{Data:} \datasource{#1}}
\newcommand{\datafn}[1]{\footnote{\datanote{#1}.}}
\newcommand{\figdir}[1]{\textit{Files:} \datasource{papers/part1/#1}}
\ifdefined\review
\newcommand{\selfcite}[1]{\cite{#1Anon}}
\else
\newcommand{\selfcite}[1]{\cite{#1}}
\fi

\newcommand{\parhead}[1]{\par\addvspace{\medskipamount}\noindent{\bfseries #1}\enspace\ignorespaces}

\OpenFOAMversions{\OF v2512}

\ifdefined\review
\Repository{}
\else
\Repository{https://github.com/isoAdvector/fluxinator}
\fi

\DOI{TBD} %leave for submission

\ifdefined\arxivcopy
\usepackage{etoolbox}
\makeatletter
\let\publname\undefined
\patchcmd{\@maketitle}{\item[DOI] \href{https://doi.org/\theDOI}{\@DOI}\\}{}{}{\PackageError{paper1}{arXiv title patch failed}{}}
\makeatother
\fi

\ifdefined\review
\fi

\begin{document}

%-----------------------------------------------------------------------
%      Manuscript title and author description
%-----------------------------------------------------------------------

%    \title[short text for running head]{full title}
\title[Consistent interfacial flow numerics -- Part I]{Consistent interfacial flow numerics -- Part I: The gravity force term}

%    Author information - hide in review mode
\ifdefined\review
  \author{}
  \address{}
  \email{}
\else
%    Author names should be given as firstname followed by surname.
%    Indicate the corresponding author with an asterisk.
%    You can optionally add an Orcid link: \orcidlink{0000-0000-0000-0000}
\author[J. Roenby]{Johan Roenby\,\orcidlink{0000-0002-7254-0546}$^{1,2,*}$}
\author[K. Missios]{Konstantinos Missios\,\orcidlink{0000-0003-2178-1718}$^2$}
\author[K. Møller]{Kasper Møller$^2$}
\author[S. Aliyar]{Sithik Aliyar\,\orcidlink{0000-0003-0188-4613}$^3$}
\author[H. Scheufler]{Henning Scheufler\,\orcidlink{0000-0003-1578-9772}$^4$}
\address{$^1$STROMNING ApS, Luftmarinegade 62, 1432 K{\o}benhavn K, Denmark}
\email{johan@stromning.com}
\address{$^2$Department of Science and Environment, Roskilde University, Universitetsvej 1,
  4000 Roskilde, Denmark}
\address{$^3$Department of Wind and Energy Systems, Technical University of Denmark,
  2800 Kgs.\ Lyngby, Denmark}
\address{$^4$DLR German Aerospace Center, Institute of Space Systems, 28359 Bremen, Germany}
\fi

%-----------------------------------------------------------------------
%      Manuscript abstract
%-----------------------------------------------------------------------

\begin{abstract}
The \OF solver \texttt{interFoam} for incompressible two-phase flows is built on the
single-phase PISO/PIMPLE algorithm. An interface advection step is added before it in the time
loop, and gravity and surface tension forces are added to the momentum equation. Otherwise it uses the
standard finite volume discretisation, which assumes that the fields vary smoothly enough to be
Taylor expanded. Across a water--air interface, however, the density jumps by a factor of about
1000. The geometric Volume-of-Fluid method isoAdvector, used in the solver
\texttt{interIsoFoam}, made the advection step consistent with a sharp interface, but it did not
remove the spurious velocities that appear near the interface, because the momentum equation is
still discretised as if the density were smooth. In this series we revisit the terms of the
momentum equation one at a time, beginning here with gravity. Both solvers absorb the gradient
part of the gravity force into an auxiliary pressure, leaving a gravity term that acts only at
the interface and depends on the local interface height. In place of this height,
\texttt{interFoam} and \texttt{interIsoFoam} use the height of the cell faces. We show that this
substitution is a key contributor to the spurious velocities in the light fluid, and that
their magnitude is not reduced by mesh refinement. Taking the height from the reconstructed
interface instead allows an exact discrete balance between pressure and gravity for a
horizontal interface on any mesh, provided that the pressure and the density use the same face
gradient and that the interface height also reaches the neighbouring cells involved in the
density gradient. We implement this geometric gravity model in a new solver called
\texttt{interIsoGoam}. In a tilted box of still water, where the original solvers drive
spurious air flows of metres per second, it keeps the velocity of the order of $10^{-9}$\,m/s, and
its advantage persists for density ratios up to $10^7$. With a hydrostatic initial pressure it
achieves the same on skewed and locally refined meshes. For moving interfaces such as waves the gain is
smaller but still of great practical value. In a progressive wave, \texttt{interIsoGoam} has the lowest maximum velocity of
the three solvers in every run. At maximum Courant numbers of 0.1 and below it therefore needs up to
43\,\% fewer steps and up to 38\,\% less wall-clock time, and its wave height error
decreases steadily as the mesh is refined. At a maximum Courant number of 0.5, however, the
wave grows by up to 12\,\% over 15 periods, so we recommend 0.1 or less. One error remains, namely an overprediction of the horizontal water velocity below the wave crest.
All three solvers exhibit this error, which grows steadily with time.
In Part~II we trace this error to the implicit discretisation of the momentum convection term.
\end{abstract}

%    No date footnote (with the corresponding author) in the review version
\ifdefined\review\date{}\else\date{\today}\fi

\dedicatory{}

\maketitle

%   Line numbering should be enable for review and disabled for final version
\ifdefined\arxivcopy\else
  \linenumbers
\fi

\section{Introduction}
\label{sec:introduction}

The Volume-of-Fluid (VoF) method \cite{Hirt1981} represents the interface between two
immiscible fluids by a volume fraction field. It solves an advection equation for this field
along with a single set of momentum and continuity equations in which the density jumps
across the interface. In \OF
\cite{Weller1998} this approach is implemented for two incompressible, immiscible fluids in the solver
\texttt{interFoam}, which is used for a wide range of interfacial flows, such as droplet impact, jet atomisation and rising bubbles
\cite{Deshpande2012,Gamet2020}. Extended with wave generation and absorption
\cite{Jacobsen2012,Higuera2013}, it is widely used for wave propagation and wave loads on structures.

At the density ratio of water and air,
about 1000, such simulations are known to be prone to numerical errors. The wave
height may grow or decay, the interface may develop unphysical undulations, and the velocity in the
crest is overpredicted \cite{Wroniszewski2014,Larsen2019}. A common feature is a large
velocity in the air just above the interface. The
air above a crest can move two to three times faster than the water in it \cite{Roenby2017}, and,
since the time step is usually limited by the largest Courant number in the domain, these air
velocities increase the cost of the simulation. Such velocities arise even for still water in a
tilted box of $1\times1$\,m, where \texttt{interFoam} and \texttt{interIsoFoam}\footnote{\texttt{interIsoFoam} is essentially
\texttt{interFoam} with the geometric isoAdvector method \cite{Roenby2016} instead of the
algebraic MULES scheme for the interface advection.} produce air velocities of
metres per second. Qwist and Christensen \cite{Qwist2023} avoid them with their so-called Direct Surface
Description method that removes the air phase from the computation.

Velocities that are caused by discretisation errors rather than by the physics are
called spurious velocities or spurious currents. They are well known in surface-tension-driven flows, where a
consistent discretisation of the force and the pressure gradient
\cite{Francois2006} combined with an accurate curvature \cite{Popinet2009} removes them
for a droplet at rest \cite{Popinet2018}. For gravity, several remedies have been proposed.
Wemmenhove et al.\ \cite{Wemmenhove2015} average the density near the free surface in a
manner consistent with gravity, and Vuk\v{c}evi\'c et al.\ \cite{Vukcevic2017} impose the
pressure jump at the interface with the ghost fluid method \cite{Fedkiw1999,Kang2000}. Mencinger and \v{Z}un
\cite{Mencinger2007} analysed the balance of discontinuous body forces on collocated grids,
and Liu et al.\ \cite{Liu2025} showed that the non-orthogonal correction of unstructured
finite volume methods creates a force imbalance at the interface.

Several codes and papers absorb the gradient part of the gravity force, $\nabla(\rho\,\grav\cdot\x)$,
with density $\rho$, gravitational acceleration $\grav$ and position $\x$, into an auxiliary
pressure. The remaining gravity term is proportional to the density gradient
and acts only at the interface, and its value depends on the height at which it is evaluated. Montazeri et al.\ \cite{Montazeri2012} showed for a level-set method that the
height must be taken at the interface, not at the position of the smeared force, since otherwise
the pressure jump depends on the interface thickness. The reduced-gravity formulation of the
Basilisk code \cite{Basilisk} evaluates the potential at the interface position in each
interface cell, obtained from height functions or from the centroid of the reconstructed
interface.

The \OF solvers \texttt{interFoam} and its geometric VoF counterpart \texttt{interIsoFoam}
also absorb the gradient part into an auxiliary pressure, $p_{rgh}$. The
remaining gravity term, $-(\grav\cdot\x)\nabla\rho$, acts only at the interface, where it
makes $p_{rgh}$ jump by an amount proportional to the density jump and to $\grav\cdot\x$ at
the height where the density changes. Both solvers take this height at the cell face centres. M{\o}ller
\cite{Moller2021} showed that the face-centre height prevents a discrete hydrostatic balance in
\texttt{interIsoFoam}, and that the geometric gravity model \texttt{gravityRecon}, which
Scheufler introduced in his PhD work \cite{Scheufler2026}, reduced the velocity after the first
time step in a tilted box by four orders of magnitude. We show that the face-centre height is
the cause of the spurious velocities in the tilted box, and that it produces a force
discretisation error that does not decrease with mesh refinement. The geometric VoF method
isoAdvector \cite{Roenby2016,Scheufler2019} reconstructs a planar interface segment in every
interface cell. We present a geometric gravity model that builds on \texttt{gravityRecon},
implemented in the solver \texttt{interIsoGoam}. It takes the interface height from the
centroids of the reconstructed segments, computes it in a simpler way than
\texttt{gravityRecon}, and assigns it to further layers of cells around the interface, which
keeps the balance with the non-orthogonal correction. For a horizontal interface the discrete gravity force is then an exact
discrete gradient, which the pressure balances on any mesh, provided that the pressure and
the density use the same face gradient and that the interface height reaches every cell in
the stencil of this gradient. In a progressive wave the model reduces the spurious
velocities in the air and the number of time steps, but not the overprediction of the crest
velocity.

This paper is the first of a series on consistent numerics for interfacial
flows in \OF, with emphasis on
geometric VoF solvers at high density ratios. It treats the gravity term only. Part~II \selfcite{RoenbyPart2}
treats the momentum convection term, which causes the overprediction of the crest
velocity that remains in progressive waves after the correction presented here
(Section~\ref{sec:wave}).

\begin{figure}[tbp]
\centering
\includegraphics[width=\linewidth]{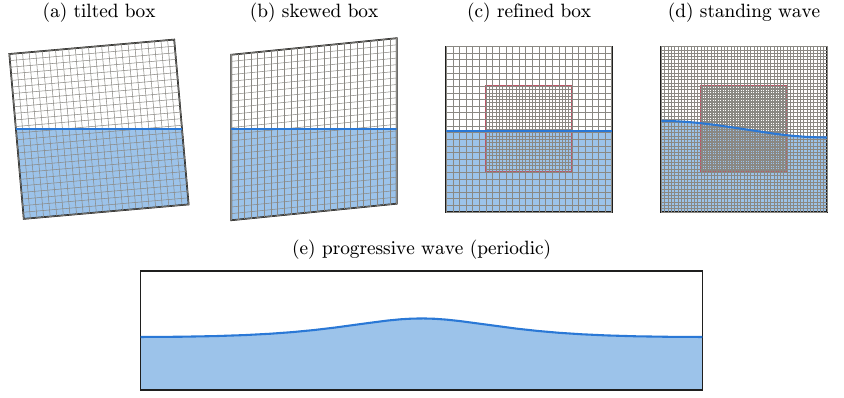}
\caption{The test cases, with the mesh and the initial interface (water shaded). (a) Horizontal
interface in a tilted box (Section~\ref{sec:tiltedBox}). (b) Skewed box and (c) box with a
refined region with a dashed outline (Section~\ref{sec:nonOrth}). (d) Standing wave across the same
refinement (Section~\ref{sec:standingWave}). (e) Progressive wave in a periodic domain one
wavelength long (Section~\ref{sec:wave}), whose meshes are too fine to draw.
\figdir{fig_cases}.}
\label{fig:cases}
\end{figure}

Section~\ref{sec:theory} analyses the discrete balance of the gravity term and defines the
new gravity model. The test cases (Fig.~\ref{fig:cases}) go from a horizontal interface at rest,
for which the exact solution is known and any velocity is spurious, to standing and
progressive waves, whose velocity fields are known from wave theory.
Sections~\ref{sec:tiltedBox} and~\ref{sec:nonOrth} test the model for a horizontal interface at
rest, on a tilted mesh and on non-orthogonal meshes (Fig.~\ref{fig:cases}a--c).
Section~\ref{sec:standingWave} considers a standing wave whose interface crosses a mesh
refinement (Fig.~\ref{fig:cases}d), and Section~\ref{sec:wave} a progressive wave
(Fig.~\ref{fig:cases}e). Section~\ref{sec:conclusion}
summarises the results and gives recommendations for users. Appendix~\ref{app:ddtCorr}
examines the \texttt{ddtCorr} term of \OF and its effect on the damping of progressive waves.

The caption of each figure and table names a directory in the code repository of this paper
(\emph{Files}). It holds the script that makes the figure or table from the runs, and the
numbers it shows. Footnotes marked \emph{Data} give the file behind a number in the text.
Together they tie each result to the numerical experiment that produced it.
The repository also holds the source of the solver \texttt{fluxinator}, which has the additions of
Part~II as defaults. \texttt{interIsoGoam} is built from the same source but keeps the defaults
of \texttt{interIsoFoam}, so it reproduces the results of this paper as released.
The results were computed with \OF v2512, and the code builds with \OF v2306 to v2606.

\section{The gravity term and its discrete balance}
\label{sec:theory}

\subsection{Governing equations and the auxiliary pressure}

We consider two incompressible, immiscible fluids with densities $\rho_1$ and $\rho_2$.
The flow is governed by
\begin{equation}
  \frac{\partial (\rho\U)}{\partial t} + \nabla\cdot(\rho\U\U)
  = -\nabla p + \nabla\cdot\boldsymbol{\tau} + \rho\grav ,
  \qquad
  \frac{\partial \rho}{\partial t} + \nabla\cdot(\rho\U) = 0 ,
  \qquad \nabla\cdot\U = 0 ,
  \label{eq:governing}
\end{equation}
where $\rho$ is the density, $\U$ the velocity, $t$ the time, $p$ the pressure,
$\boldsymbol{\tau}$ the viscous stress tensor and $\grav$ the gravitational acceleration.
In all test cases of this paper both fluids are inviscid, so that $\boldsymbol{\tau} = 0$.
The density is $\rho_1$ in fluid 1 and $\rho_2$ in fluid 2, and fluid 1 is the heavier
fluid, $\rho_1 > \rho_2$ (water and air in the test cases). In the VoF method each cell
carries the volume fraction $\alpha$ of fluid 1 in the cell, which the advection step updates
from the volume of fluid 1 transported through the cell faces, and the cell density is
$\rho = \alpha\rho_1 + (1-\alpha)\rho_2$. Surface tension is not considered in this paper. The solvers \texttt{interFoam} and \texttt{interIsoFoam} split the pressure as
$p = p_{rgh} + \rho\,gh$, where $gh = \grav\cdot\x$ is the height scaled by gravity, with
the position $\x$ measured from a reference level \cite{Rusche2002}. This removes the large hydrostatic
pressure gradient from the pressure variable and simplifies the pressure boundary conditions. Since
\begin{equation}
  \nabla(\rho\,\grav\cdot\x) = \rho\grav + (\grav\cdot\x)\,\nabla\rho ,
  \label{eq:gradIdentity}
\end{equation}
it follows that
\begin{equation}
  -\nabla p + \rho\grav = -\nabla p_{rgh} - (\grav\cdot\x)\,\nabla\rho .
  \label{eq:prgh}
\end{equation}
The remaining gravity term $-(\grav\cdot\x)\nabla\rho$ thus acts only where the density varies, that is, at the interface, and
it is proportional to the height at which the density changes.
A full description of the solvers \texttt{interFoam} and \texttt{interIsoFoam} is beyond
the scope of this paper. For \texttt{interFoam} we refer the reader to
\cite{Deshpande2012,Qwist2023}, and for \texttt{interIsoFoam} to \cite{Roenby2016,Roenby2019}.

\subsection{The face-based gravity term and the discrete balance}

The solvers evaluate the right-hand side of Eqn.~\ref{eq:prgh} on the cell faces. For a face
$f$ between the owner cell $P$ and the neighbour cell $N$, the gravity term of \texttt{interFoam}
and \texttt{interIsoFoam} is
\begin{equation}
  F_{g,f} = -\,(\grav\cdot\x_f)\,\frac{\rho_N - \rho_P}{|\mathbf{d}_f|},
  \label{eq:basic}
\end{equation}
where $\x_f$ is the face centre and $\mathbf{d}_f$ the vector between the cell centres.
We call this the \emph{basic} model. In \texttt{interFoam} this is the natural choice, because its algebraic VoF method
smears the interface over a few cells and provides no interface position. The height is therefore
taken where the density difference occurs, at the face. The solver \texttt{interIsoFoam} was derived from
\texttt{interFoam} with a geometric advection step and inherited the gravity term unchanged, although
its reconstruction provides the interface position. 

The face flux is
\begin{equation}
  \phi_f = \phi^*_f + r_{A,f}\,|\mathbf{S}_f|\,\bigl(F_{g,f} - \nabla_f p_{rgh}\bigr),
  \label{eq:faceFlux}
\end{equation}
where $\phi^*_f$ contains the inertial and viscous contributions, $\mathbf{S}_f$ is the
face area vector with unit normal $\mathbf{n}_f = \mathbf{S}_f/|\mathbf{S}_f|$, $\nabla_f$ the face-normal gradient and $r_{A,f}$ the interpolated
reciprocal of the diagonal of the momentum matrix. The pressure $p_{rgh}$ is determined
such that $\phi$ is divergence free. On a non-orthogonal mesh, $1/|\mathbf{d}_f|$ is
replaced by the delta coefficient $1/(\mathbf{n}_f\cdot\mathbf{d}_f)$ of the face, and the face-normal gradients of $\rho$ and
$p_{rgh}$ receive a non-orthogonal correction. We ignore these corrections here and return to
them in Section~\ref{sec:nonOrth}.

\begin{figure}[tbp]
\centering
\includegraphics[width=\linewidth]{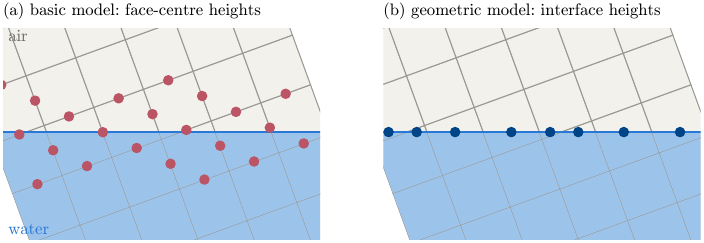}
\caption{The height in the gravity term on a tilted mesh with a horizontal interface (tilt
exaggerated). (a) The basic model uses the centres of the faces across which the density
changes (circles), which lie at different heights. (b) The geometric gravity model uses the centroids
of the reconstructed interface segments (circles), which all lie at the interface height.
\figdir{fig_height}.}
\label{fig:height}
\end{figure}

For a fluid at rest, $\phi^* = 0$, and the rest state persists only if there is a cell
field $p_{rgh}$ with
\begin{equation}
  \frac{p_{rgh,N} - p_{rgh,P}}{|\mathbf{d}_f|} = F_{g,f}
  \quad\text{on every internal face,}
  \label{eq:balance}
\end{equation}
that is, if the face values $F_{g,f}$ form a discrete gradient.
To test this, consider a closed loop $\mathcal{L}$ of cells
$C_0, C_1, \dots, C_m = C_0$ in which consecutive cells $C_{k-1}$ and $C_k$ are \emph{face neighbours},
that is, they share the face $f_k$. Let $s_k = +1$ if the loop crosses $f_k$ from owner to neighbour and $s_k = -1$
otherwise, so that for any cell field $\psi$ the difference along the loop is
\begin{equation}
  \Delta_k\psi = \psi_{C_k} - \psi_{C_{k-1}} = s_k\,(\psi_N - \psi_P) .
  \label{eq:loopDifference}
\end{equation}
The loop sum of the
gravity term is
\begin{equation}
  \Gamma_{\mathcal{L}} = \sum_{k=1}^{m} s_k\,|\mathbf{d}_{f_k}|\,F_{g,f_k} .
  \label{eq:loopSum}
\end{equation}
If Eqn.~\ref{eq:balance} holds, every term is a difference of $p_{rgh}$ along the loop,
and the sum telescopes:
\begin{equation}
  \Gamma_{\mathcal{L}} = \sum_{k=1}^{m} \Delta_k p_{rgh}
  = p_{rgh,C_m} - p_{rgh,C_0} = 0 ,
\end{equation}
the discrete counterpart of $\oint \nabla p\cdot\mathrm{d}\mathbf{l} = 0$. Conversely,
if $\Gamma_{\mathcal{L}} = 0$ for every loop, a field $p_{rgh}$ that satisfies
Eqn.~\ref{eq:balance} is obtained by summing $|\mathbf{d}_f|F_{g,f}$ along any path from a
reference cell, since the result does not depend on the path. The rest state is therefore
discretely possible if and only if all loop sums vanish. For the basic model
(Eqn.~\ref{eq:basic}),
\begin{equation}
  \Gamma_{\mathcal{L}} = -\sum_{k=1}^{m}
  (\grav\cdot\x_{f_k})\,\Delta_k\rho .
  \label{eq:loopImbalance}
\end{equation}
Since $\sum_k \Delta_k\rho = 0$ by the same telescoping,
$\Gamma_{\mathcal{L}}$ vanishes when all faces that carry a density difference lie at the
same height.
It also vanishes for a horizontal interface on a mesh of horizontal rows of cells. The
density then differs only between rows, so only the faces between two rows carry a density
difference, and these faces all have the same height and the same density difference. A
closed loop crosses each such boundary as often upwards as downwards, so the contributions
cancel.
On a mesh that is not aligned with gravity, the faces that carry a density difference lie at
different heights,
so that in general $\Gamma_{\mathcal{L}}\neq 0$, and the part of the force that is not a gradient drives a
flux that the pressure cannot remove. On a tilted mesh, the faces across which the
density changes have centres at many different heights, both above and below the interface.
Figure~\ref{fig:height}a illustrates this. The result is a spurious velocity in a fluid that should be at rest.

For a sloping interface the remaining gravity term $-(\grav\cdot\x)\nabla\rho$ is itself
not a gradient, since its curl, $\nabla\rho\times\grav$, is non-zero wherever the interface is not
horizontal. The pressure cannot balance such a force, and the unbalanced part drives the motion
of the interface. The discrete loop sums (Eqn.~\ref{eq:loopSum}) should then reproduce the circulation of
this term, not vanish. The tests of Sections~\ref{sec:tiltedBox}
and~\ref{sec:nonOrth} use a horizontal interface at rest, for which the exact loop sums are zero, while
Sections~\ref{sec:standingWave} and~\ref{sec:wave} use moving, sloping interfaces.

\subsection{The height at the interface}
\label{sec:heightAtInterface}

For a sharp interface $I$ with unit normal $\mathbf{n}_I$ pointing into fluid 1, the
density gradient is a surface delta function, $\nabla\rho = (\rho_1-\rho_2)\,\mathbf{n}_I\,
\delta_I$. Averaging the face-normal component of the gravity term over a control volume
$V_f$ that extends from the centre of cell $P$ to that of cell $N$ gives
\begin{equation}
  \mathbf{n}_f\cdot\frac{1}{V_f}\int_{V_f} -(\grav\cdot\x)\,\nabla\rho\,\mathrm{d}V
  = -\frac{\rho_1-\rho_2}{V_f}\int_{I\cap V_f} (\grav\cdot\x)\,
    (\mathbf{n}_I\cdot\mathbf{n}_f)\,\mathrm{d}S .
  \label{eq:faceAverage}
\end{equation}
Comparing with the form $-gh_f\,(\rho_N-\rho_P)/|\mathbf{d}_f|$ of the discrete term, the
height $gh_f$ should be the weighted mean of $\grav\cdot\x$ over the part of the interface
in $V_f$,
\begin{equation}
  gh_f \approx \frac{\int_{I\cap V_f} (\grav\cdot\x)\,(\mathbf{n}_I\cdot\mathbf{n}_f)\,\mathrm{d}S}
                    {\int_{I\cap V_f} (\mathbf{n}_I\cdot\mathbf{n}_f)\,\mathrm{d}S} .
  \label{eq:ghfInterface}
\end{equation}
Since $\grav\cdot\x$ is linear, its mean over a planar interface segment is its value at
the centroid of the segment. A geometric VoF method provides these centroids
(Fig.~\ref{fig:height}b). The same principle underlies the \emph{effective elevation} of
Montazeri et al.\ \cite{Montazeri2012}, who took the height at the closest point of the zero
level set. With a geometric VoF method the height is available directly from the
reconstruction. Basilisk \cite{Basilisk}, which also uses the centroid
(Section~\ref{sec:introduction}), extends the potential at the refinement boundaries of its
tree grid to every face that the gradient of the volume fraction reaches. The layers introduced in Section~\ref{sec:geometricModel} serve the same purpose on general
meshes, including the wider stencil of the non-orthogonal correction
(Section~\ref{sec:nonOrth}).

\parhead{Error of the basic model.} The basic model uses $\grav\cdot\x_f$ instead of
Eqn.~\ref{eq:ghfInterface}. The two differ by an amount of order $|\grav|\,\Delta x$ for a cell size $\Delta x$, while the
density difference across an interface face is of order $|\rho_1-\rho_2|$ over the
distance $\Delta x$. The discretisation error of the force per unit volume is therefore of order
$|\rho_1-\rho_2|\,|\grav|$, independent of the cell size, and in the light fluid 2 it
corresponds to an acceleration of order $(|\rho_1-\rho_2|/\rho_2)\,|\grav|$. The pressure
removes the part of this error that is a discrete gradient. The part that remains is much
smaller than this estimate. In the tilted box of Section~\ref{sec:tiltedBox} the
acceleration in the first time step is 3.4 to $5.1\,|\grav|$ (Tab.~\ref{tab:firstStep})
instead of about $1000\,|\grav|$. Like the estimate, however, it does not decrease with the cell size.

\parhead{Exact balance with the interface height.} Without surface tension the pressure
$p$ is continuous across the interface, so $p_{rgh}$ jumps by
$-(\rho_1-\rho_2)\,(\grav\cdot\x_I)$ at an interface point $\x_I$. With the interface
height in the face term,
\begin{equation}
  F_{g,f} = -\,(\grav\cdot\x_I)\,\frac{\rho_N - \rho_P}{|\mathbf{d}_f|}
          = \frac{\bigl(-\rho\,\grav\cdot\x_I\bigr)_N - \bigl(-\rho\,\grav\cdot\x_I\bigr)_P}{|\mathbf{d}_f|} .
  \label{eq:exact}
\end{equation}
For a horizontal interface, $\grav\cdot\x_I$ is a constant, so Eqn.~\ref{eq:exact} is the discrete
gradient of the cell field $-\rho\,\grav\cdot\x_I$ on any mesh, the balance
(Eqn.~\ref{eq:balance}) holds, and the rest state is preserved. On a non-orthogonal mesh
this requires that the same face-normal gradient, including its correction, is used for
$p_{rgh}$ and $\rho$, and that every cell entering the gradient of $\rho$ carries the interface
height (Section~\ref{sec:nonOrth}).

\subsection{The geometric gravity model}
\label{sec:geometricModel}

The solver \texttt{interIsoGoam} is \texttt{interIsoFoam} with the gravity term replaced by a run-time selectable
model. The model is selected in the file \texttt{system/fvSchemes} of the case (for example
\datasource{run/tiltedBox/baseCase/system/fvSchemes}) by
\begingroup\small
\begin{verbatim}
interfaceForces { gravity { model geometric; } }
\end{verbatim}
\endgroup
\noindent Without this entry \texttt{interIsoGoam} uses the basic model. Besides the basic model it provides:

\parhead{\texttt{constantHeight}.} This model sets $\grav\cdot\x_f = \grav\cdot\x_I$ on all faces, with $\x_I$ from
the initial condition. It applies only to a horizontal interface at rest, and its sole
purpose is to serve as the exact
reference in Sections~\ref{sec:tiltedBox} and~\ref{sec:nonOrth}. It should not be used in dynamic cases.

\begin{figure}[tbp]
\centering
\includegraphics[width=0.45\linewidth]{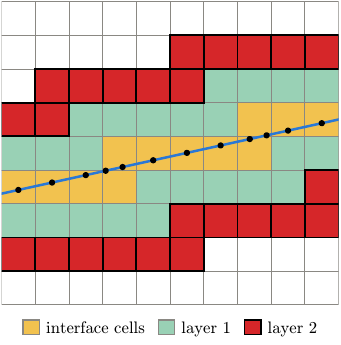}
\caption{The cells that receive the interface height in the geometric gravity model, for a
planar interface (blue line) on a square mesh: interface cells, layer 1 (point neighbours of
interface cells) and layer 2 (point neighbours of layer 1). The dots are the centroids of the interface
segments.
\figdir{fig_layers}.}
\label{fig:layers}
\end{figure}

\parhead{\texttt{geometric}.} After the isoAdvector advection step the interface is reconstructed
\cite{Scheufler2019}, giving in every interface cell $i$ a planar segment with centroid
$\x_{c,i}$ and unit normal $\hat{\mathbf{n}}_i$. An interface cell is a cell with
$\epsilon < \alpha < 1-\epsilon$, where $\epsilon$ is a small tolerance (\texttt{surfCellTol},
$10^{-9}$ to $10^{-6}$ in this paper). The \emph{point neighbours} of a cell are the cells that share at least one mesh point with it.
A cell field $gh$ is defined by
\begin{equation}
  gh_i =
  \begin{cases}
    \grav\cdot\x_{c,i} & \text{interface cells,}\\[4pt]
    \grav\cdot\dfrac{\sum_j w_{ij}\,\x_{c,j}}{\sum_j w_{ij}},
    \quad w_{ij} = \bigl(\hat{\mathbf{d}}_{ij}\cdot\hat{\mathbf{n}}_j\bigr)^2
      & \text{layer 1,}\\[10pt]
    \dfrac{1}{|\mathcal{A}_i|}\displaystyle\sum_{j\in\mathcal{A}_i} gh_j
      & \text{layers } 2,\dots,n_L,\\[10pt]
    \grav\cdot\x_{C,i} & \text{all other cells,}
  \end{cases}
  \label{eq:geometric}
\end{equation}
where layer 1 consists of the point neighbours of interface cells that are not interface
cells themselves, $j$ runs
over the interface cells among the point neighbours of cell $i$, $\hat{\mathbf{d}}_{ij}$ is
the unit vector from the centre of cell $i$ to $\x_{c,j}$, and $\x_{C,i}$ is the cell
centre. Layer $k\ge 2$ consists of the point neighbours of layer $k-1$ not in a lower layer,
and $\mathcal{A}_i$ is the set of point neighbours of cell $i$ that are interface cells or
lie in a lower layer, with $|\mathcal{A}_i|$ the number of cells in it (Fig.~\ref{fig:layers}).  The number of layers $n_L$ is a run-time parameter (\texttt{nLayers}, next to \texttt{model}, default 2).
In our tests the second layer increased the time per step by less than
2\,\%.\datafn{papers/part1/fig_waveCost/waveCost.txt}
The face value is interpolated from the cells,
$gh_f = w_f\,gh_P + (1-w_f)\,gh_N$ with the linear interpolation weight $w_f$, so that it
equals the interface height where both cells of the face carry it, and the face term is
$F_{g,f} = -\,gh_f\,(\rho_N-\rho_P)/|\mathbf{d}_f|$. An interface that lies exactly on mesh
faces leaves no interface cells. A face between a full and an empty cell is then treated as
an interface segment for both adjacent cells, with the face centre as its centroid and the
face normal as its normal.

\parhead{Cells that need the interface height.} The balance requires
$gh_f = \grav\cdot\x_I$ on every face with a non-zero density gradient, and since $gh_f$
is interpolated, both cells of such a face must carry the interface value. With a
geometric VoF method and the two-point gradient of Eqn.~\ref{eq:basic}, every such face has an
interface cell on at least one side and an interface cell or a layer-1 cell on the other.
For $n_L \ge 1$ the geometric gravity model therefore reproduces Eqn.~\ref{eq:exact} for a horizontal
interface, to the accuracy with which the reconstruction reproduces a plane. With
$n_L = 0$ the faces between interface cells and their face neighbours receive the mean of an
interface height and a cell-centre height, and the balance is lost
(Section~\ref{sec:tiltedBox}). The non-orthogonal correction widens the stencil of the
density gradient and can require $n_L = 2$ (Section~\ref{sec:nonOrth}). The same reasoning
explains why the model requires a geometric VoF method, since with an algebraic method the
density varies over several cells.

The model replaces only the face gravity term, and the pressure equation is unchanged. The
reconstruction is the one at the end of the advection step, so the gravity term uses the
interface at the new time level, consistent with the new density. Table~\ref{tab:solvers}
summarises the solvers compared.

\begin{table}[tbp]
\centering
\caption{The solvers compared. With the basic model and one outer corrector,
\texttt{interIsoGoam} is identical to \texttt{interIsoFoam}. The \texttt{interIsoFoam}
results of Sections~\ref{sec:tiltedBox} and~\ref{sec:nonOrth} are computed in this way, those
of the waves with \texttt{interIsoFoam} of \OF.
\textit{Source:} \datasource{applications/solvers/interIsoGoam} and \datasource{src/gravityModels}.}
\label{tab:solvers}

\small
\begin{tabular}{@{}lccc@{}}
\toprule
 & interFoam & interIsoFoam & interIsoGoam \\
\midrule
Interface advection & algebraic (MULES) & geometric (isoAdvector) & geometric (isoAdvector) \\
Interface reconstruction & none & isoAlpha or plicRDF \cite{Scheufler2019} & isoAlpha or plicRDF \cite{Scheufler2019} \\
Height in gravity term & $\grav\cdot\x_f$ & $\grav\cdot\x_f$ & basic, constantHeight or geometric \\
\bottomrule
\end{tabular}
\end{table}

\section{A horizontal interface at rest in a tilted box}
\label{sec:tiltedBox}

\parhead{Setup.} A two-dimensional square box of side 1\,m, discretised with $25\times25$
square cells, is rotated by $5^\circ$ about its centre, while gravity,
$\grav = (0,0,-9.81)$\,m/s$^2$ in the coordinates $(x,y,z)$, remains vertical
(Fig.~\ref{fig:cases}a). Water
($\rho_1 = 1000$\,kg/m$^3$) fills the box below $z = 0.5$\,m and air ($\rho_2 = 1$\,kg/m$^3$)
above it. The volume fractions are initialised exactly from the plane with the
\texttt{setAlphaField} utility of \OF. The walls are slip walls, and the fluids start at rest. The exact solution is a fluid at rest, so
any velocity is spurious. The case was introduced in the isoAdvector repository
\cite{isoAdvectorRepo} and used by Qwist and Christensen \cite{Qwist2023}.

All runs use a time step of $\Delta t = 0.01$\,s, Euler time stepping, one outer and three inner
pressure correctors without momentum predictor, and a pressure tolerance of $10^{-10}$
($10^{-12}$ in the final inner corrector). The interface is reconstructed with
\texttt{plicRDF}, which Scheufler and Roenby \cite{Scheufler2019}
showed to be the most accurate of the schemes they tested.
Unless stated
otherwise, the geometric gravity model uses the default $n_L = 2$.

\begin{figure}[tbp]
\centering
\includegraphics[width=\linewidth]{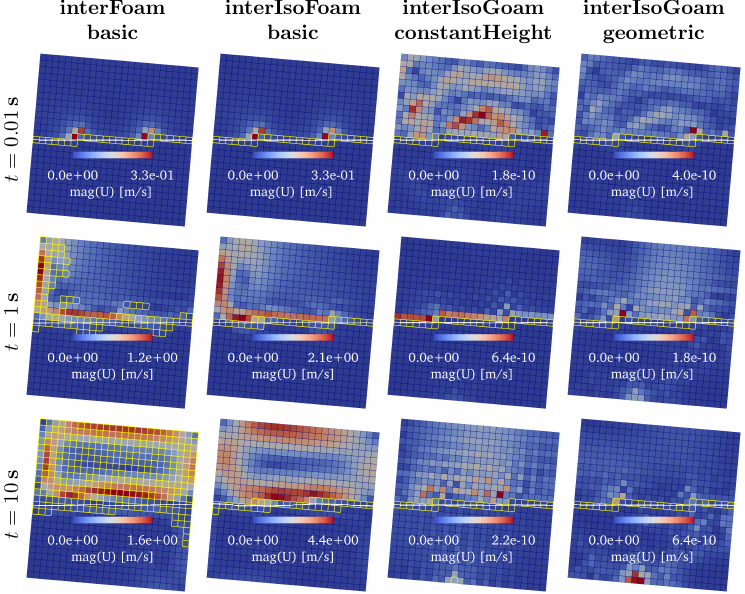}
\caption{Velocity magnitude in the tilted box after 1, 100 and 1000 time steps
($t = 0.01$, 1 and 10\,s, in rows) for each solver and gravity model (columns). Each panel has its
own colour scale, adjusted to its maximum value. Yellow outlines mark the cells with $10^{-9} < \alpha < 1-10^{-9}$, and the white line is the
contour $\alpha = 0.5$.
\figdir{fig_tiltedBoxFields}.}
\label{fig:tiltedBoxFields}
\end{figure}

\begin{table}[tbp]
\centering
\caption{Maximum velocity magnitude [m/s] in the tilted box after the first time step
($t = 0.01$\,s) and after 1000 steps ($t = 10$\,s), and its peak over the run. $n_L$ is the number
of layers that receive the interface height, Eqn.~\ref{eq:geometric}.
\figdir{tab_tiltedBox}.}
\label{tab:tiltedBox}
\small
\begin{tabular}{@{}lccc@{}}
\toprule
Configuration & $t = 0.01$\,s & $t = 10$\,s & peak, $t \le 10$\,s \\
\midrule
interFoam, basic                       & $3.3\times10^{-1}$  & $1.6$ & $1.6$ \\
interIsoFoam, basic                    & $3.3\times10^{-1}$  & $4.4$ & $4.6$ \\
interIsoGoam, constantHeight           & $1.8\times10^{-10}$ & $2.2\times10^{-10}$ & $2.0\times10^{-9}$ \\
interIsoGoam, geometric, $n_L = 0$     & $5.3\times10^{-1}$  & $5.2$ & $5.4$ \\
interIsoGoam, geometric, $n_L = 1$     & $4.0\times10^{-10}$ & $7.2\times10^{-10}$ & $2.5\times10^{-9}$ \\
interIsoGoam, geometric, $n_L = 2$     & $4.0\times10^{-10}$ & $6.4\times10^{-10}$ & $2.4\times10^{-9}$ \\
\bottomrule
\end{tabular}
\end{table}

\begin{table}[tbp]
\centering
\caption{Spurious acceleration of the basic model in the first time step of the tilted
box, in units of $|\grav|$. It is the same for \texttt{interFoam} and \texttt{interIsoFoam}
to all digits shown.
\figdir{tab_firstStep}.}
\label{tab:firstStep}
\small
\begin{tabular}{@{}lcccc@{}}
\toprule
Mesh & $25\times25$ & $50\times50$ & $100\times100$ & $200\times200$ \\
\midrule
$\max|\U|/(\Delta t\,|\grav|)$ after the first step & $3.36$ & $5.03$ & $5.04$ & $5.11$ \\
\bottomrule
\end{tabular}
\end{table}

\parhead{The basic model.} With the basic gravity term, \texttt{interFoam} and \texttt{interIsoFoam} produce
the same velocity of 0.33\,m/s in the first time step (Tab.~\ref{tab:tiltedBox}). The
interface has not moved, so this velocity is caused by the gravity term alone. It is
concentrated in a few cells next to the interface (Fig.~\ref{fig:tiltedBoxFields}, top row). Starting from rest, the first
velocity update is $\U = \Delta t\,\mathbf{F}/\rho$, with $\mathbf{F}$ the part of the
force that the pressure cannot balance. The spurious acceleration $\max|\U|/\Delta t$ is
therefore independent of the time step. It lies between $3.4$ and $5.1\,|\grav|$ (Tab.~\ref{tab:firstStep}) and does not decrease when the mesh is refined from
$25\times25$ to $200\times200$ cells, as estimated in Section~\ref{sec:heightAtInterface}.
Runs with time steps from $10^{-4}$ to 1\,s give the same acceleration to all digits shown.
After the first step the velocities grow into a circulation that fills the air
(Fig.~\ref{fig:tiltedBoxFields}, middle and bottom rows), reaching 1.6\,m/s for \texttt{interFoam} and 4.4\,m/s
for \texttt{interIsoFoam} after 10\,s (Tab.~\ref{tab:tiltedBox}).

\parhead{The interface height.} With the exact interface height (\texttt{constantHeight}), the
velocity stays below $2.0\times10^{-9}$\,m/s (Tab.~\ref{tab:tiltedBox}). We attribute this
level to the pressure solution, whose final tolerance of $10^{-12}$ limits the attainable
accuracy. The geometric gravity model does not use the known initial position. Instead it takes the
height from the interface reconstructed from the volume fraction field in every time step.
With one or two layers it stays within a factor of 3.4 of the \texttt{constantHeight} reference. Without layers ($n_L = 0$), the velocities are larger than with the basic model,
as anticipated in Section~\ref{sec:geometricModel}.

\begin{figure}[tbp]
\centering
\includegraphics[width=0.62\linewidth]{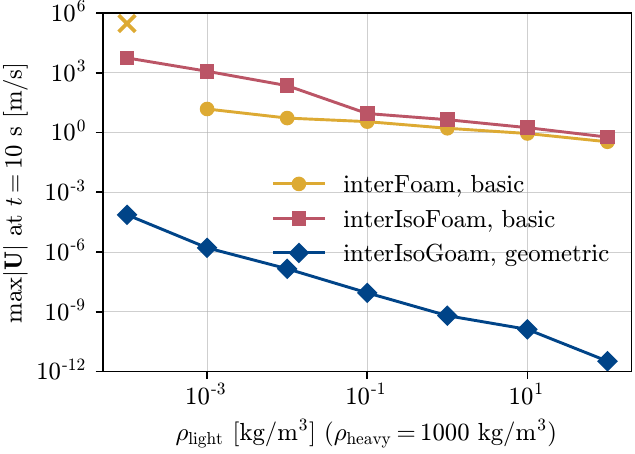}
\caption{Maximum velocity magnitude in the tilted box after 10\,s against the density of
the light fluid. The cross marks the \texttt{interFoam} run at $10^{-4}$\,kg/m$^3$, which diverged at
$t = 1.84$\,s.
\figdir{fig_tiltedBoxRhoAir}.}
\label{fig:tiltedBoxRhoAir}
\end{figure}

\parhead{Density ratio.} The spurious velocities of the basic model increase as the light
fluid becomes lighter, and \texttt{interFoam} diverges at the lowest density tested here
(Fig.~\ref{fig:tiltedBoxRhoAir}), as estimated in Section~\ref{sec:heightAtInterface}. The
force error is proportional to the density jump, which remains almost constant, and the initial
acceleration in the light fluid is inversely proportional to its density. The
geometric gravity model removes this error. Its velocities stay eight to eleven orders of
magnitude below those of \texttt{interIsoFoam} for density ratios from 10 to $10^7$, so its
advantage is not limited to water and air. Its residual velocity also increases as the light
fluid becomes lighter, by about one order of magnitude per decade of density.

\section{A horizontal interface at rest on non-orthogonal meshes}
\label{sec:nonOrth}

The meshes of engineering simulations are rarely orthogonal, and \OF corrects the face
gradients for this. This section investigates whether the balance of
Section~\ref{sec:theory} still holds with this correction. On non-orthogonal meshes the \emph{corrected} face-normal gradient and Laplacian
schemes of \OF, selected in \texttt{fvSchemes}, add a correction to the two-point difference,
\begin{equation}
  \nabla_f\psi = \frac{\psi_N-\psi_P}{\mathbf{n}_f\cdot\mathbf{d}_f}
    + \mathbf{k}_f\cdot(\nabla\psi)_f ,
  \qquad
  \mathbf{k}_f = \mathbf{n}_f - \frac{\mathbf{d}_f}{\mathbf{n}_f\cdot\mathbf{d}_f},
  \label{eq:corrected}
\end{equation}
where $(\nabla\psi)_f$ is the cell gradient, linearly interpolated from the cells $P$ and $N$.
The cell gradient is computed with the scheme selected in \texttt{gradSchemes}, here the Gauss
gradient with linear interpolation. The
\emph{uncorrected} schemes omit the second term. In the pressure equation the correction
is evaluated explicitly with the latest $p_{rgh}$, and non-orthogonal corrector loops
repeat the solution with an updated correction.

\parhead{The discrete rest state.} The balance argument of Section~\ref{sec:theory} holds
for any face gradient that is linear in the cell values. For a horizontal interface through the point $\x_I$, consider the hydrostatic cell field
\begin{equation}
  p_{rgh,i}^{\,\mathrm{hs}} = -\rho_i\,(\grav\cdot\x_I) .
  \label{eq:prghHs}
\end{equation}
Linearity gives
\begin{equation}
  \nabla_f p_{rgh}^{\,\mathrm{hs}} = -(\grav\cdot\x_I)\,\nabla_f\rho ,
  \label{eq:prghHsGrad}
\end{equation}
and with $\phi^* = 0$ in Eqn.~\ref{eq:faceFlux} the face flux is
\begin{equation}
  \phi_f = r_{A,f}\,|\mathbf{S}_f|\,\bigl(\grav\cdot\x_I - gh_f\bigr)\,\nabla_f\rho .
  \label{eq:nonOrthBalance}
\end{equation}
The rest state is therefore a discrete solution if (i) the same face gradient scheme is
used for $p_{rgh}$ and for $\rho$, and (ii) $gh_f = \grav\cdot\x_I$ on every face with
$\nabla_f\rho\neq 0$. It remains to determine whether the solver reaches this state from
its initial condition, and which faces have $\nabla_f\rho\neq 0$ when the correction is
included. 

The runs below use the fluids, boundary conditions and solver settings of
Section~\ref{sec:tiltedBox} for 10 time steps. Unless stated otherwise, the corrected
schemes are used, $p_{rgh}$ is initially zero, and no non-orthogonal corrector loops are performed. The
variations are the uncorrected schemes, $n_c$ non-orthogonal corrector loops, and a
\emph{hydrostatic start}, in which $p_{rgh}$ is initialised with
$p_{rgh}^{\,\mathrm{hs}}$~(Eqn.~\ref{eq:prghHs}) by the utility \texttt{setHydrostaticPressure}
of the code repository.

\subsection{Skewed box}

The box of Section~\ref{sec:tiltedBox} is replaced by a parallelogram whose $25\times25$ cell
rows rise by 0.1\,m over the width of 1\,m, while the columns remain vertical
(Fig.~\ref{fig:cases}b). This is the simplest non-orthogonal mesh. All faces are
non-orthogonal by $5.7^\circ$, opposite faces of every cell are parallel, and the interface at
$z = 0.5$\,m is horizontal but cuts faces at different heights. Rotating the mesh by
$90^\circ$ makes the faces that carry the density jump level without changing the
non-orthogonality. Comparing the two orientations separates the error of the face heights
from the effect of the non-orthogonal correction.

\begin{table}[tbp]
\centering
\caption{Maximum velocity magnitude [m/s] in the skewed box after 10 time steps, with
corrected and uncorrected schemes. $n_c$ is the number of non-orthogonal corrector loops in every
time step, from a zero
initial $p_{rgh}$, and \emph{hydrostatic} means a hydrostatic initial $p_{rgh}$ and $n_c = 0$. The last column is for the
same mesh rotated by $90^\circ$, so that the cell rows are level and the columns lean.
\figdir{tab_nonOrthSkew}.}
\label{tab:nonOrthSkew}
\small
\begin{tabular}{@{}lccccc@{}}
\toprule
 & \multicolumn{3}{c}{corrected} & \multicolumn{2}{c}{uncorrected} \\
\cmidrule(lr){2-4}\cmidrule(lr){5-6}
 & $n_c = 0$ & $n_c = 2$ & hydrostatic & $n_c = 0$ & rows level \\
\midrule
interFoam, basic              & 5.6                & 5.7                & 5.6                 & $7.6\times10^{-1}$  & $3.9\times10^{-10}$ \\
interIsoFoam, basic           & 7.8                & 7.8                & 7.8                 & 1.3                 & $6.0\times10^{-10}$ \\
interIsoGoam, constantHeight  & $3.6\times10^{-2}$ & $2.1\times10^{-8}$ & $1.0\times10^{-9}$  & $4.0\times10^{-11}$ & $4.6\times10^{-10}$ \\
interIsoGoam, geometric, $n_L = 2$ & $3.5\times10^{-2}$ & $2.0\times10^{-8}$ & $7.7\times10^{-10}$ & $4.1\times10^{-11}$ & $4.7\times10^{-10}$ \\
\bottomrule
\end{tabular}
\end{table}

\begin{figure}[tbp]
\centering
\includegraphics[width=\linewidth]{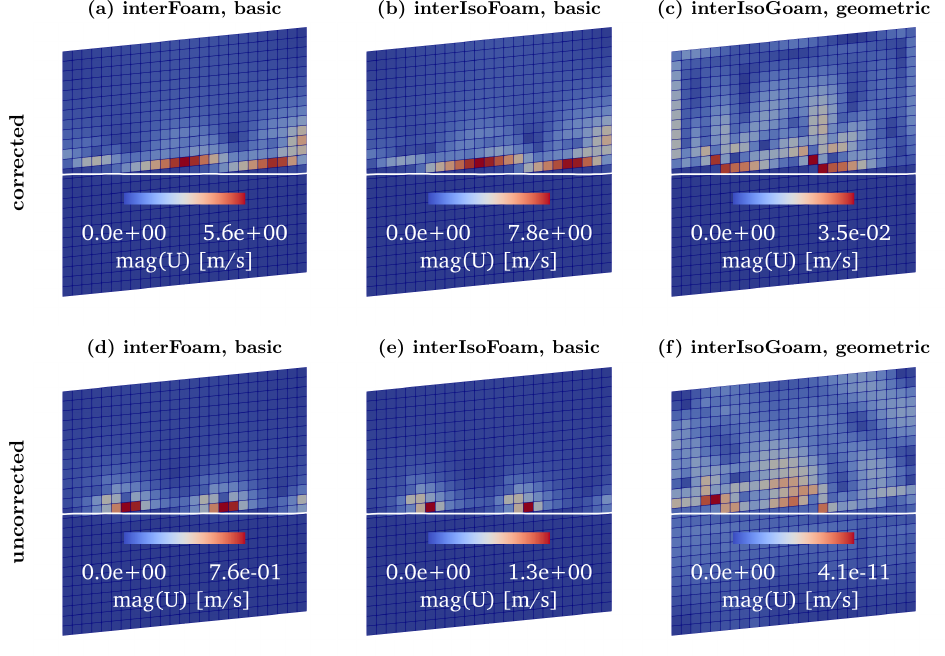}
\caption{Velocity magnitude in the skewed box after 10 time steps, with corrected (a--c) and
uncorrected (d--f) schemes, zero initial $p_{rgh}$ and $n_c = 0$. Each panel has its own
colour scale, adjusted to its maximum value, and the white line is the contour $\alpha = 0.5$.
\figdir{fig_skewedBox}.}
\label{fig:skewedBox}
\end{figure}

Neither the geometric gravity model
alone nor the uncorrected schemes alone keep the fluid at rest (Tab.~\ref{tab:nonOrthSkew},
Fig.~\ref{fig:skewedBox}).

 With the basic model the velocities are of order
1--8\,m/s with either scheme. With uncorrected schemes this is the error of
Section~\ref{sec:theory}. When the mesh is rotated so that the faces carrying the density
jump are level, the basic model also keeps the fluid at rest
(Tab.~\ref{tab:nonOrthSkew}, last column). With corrected
schemes, a zero initial $p_{rgh}$ and $n_c = 0$, the geometric gravity model and the exact
interface height (\texttt{constantHeight}) give the same velocity, $3.5\times10^{-2}$\,m/s, so
this error is not caused by an inaccurate interface height. It is a start-up error that arises because in the first pressure solution the correction of the $p_{rgh}$
gradient is evaluated with $p_{rgh} = 0$, while the correction of the density gradient in
the gravity term is evaluated with the actual density. A hydrostatic start eliminates this
error. Non-orthogonal corrector loops in every time step reduce it by several orders of
magnitude (Tab.~\ref{tab:nonOrthSkew}), but more than two loops give little further
reduction.\datafn{run/skewedBox/data/nonOrthCorrectors_10steps.txt} Neither remedy reduces
the velocities of the basic model. Liu et al.\ \cite{Liu2025} analysed this imbalance of the explicit non-orthogonal correction
for a droplet and a water column at rest. They repeated the correction until the initial
residual of the pressure equation was below the solver tolerance and, with a prescribed exact
interface height in the gravity term, kept the water column at rest to about
$10^{-8}$\,m/s on three-dimensional unstructured meshes with non-orthogonality of up to about
$60^\circ$. A prescribed height is available only for an interface at rest. The geometric
gravity model takes the height from the reconstructed interface, also when the interface moves. The
next subsection shows that on a locally refined mesh this height must also reach a second layer
of cells.

\subsection{Box with a refined region}

\parhead{Setup.} Mesh generators
such as \texttt{snappyHexMesh} refine a hexahedral base mesh locally and snap it to the geometry,
which produces polyhedral and non-orthogonal cells. Here we consider one source of such cells,
the boundary of a refined region. There,
a coarse cell face is split into fine faces whose centres lie at different heights, and the
cells no longer have parallel opposite faces. This case tests the model where the interface
crosses such a boundary. The mesh is the box of Section~\ref{sec:tiltedBox}, now untilted, but with
the central $13\times13$ cells ($0.24 \le x, z \le 0.76$\,m) split into $2\times2$ cells
(Fig.~\ref{fig:cases}c). At the refinement boundary each coarse
face is split into two faces whose centres lie a quarter of a coarse cell above and below the
coarse cell centre, and the interface crosses this boundary at the left and right sides of
the refined region. The still water level is $z_w = 0.49$\,m, a quarter of a coarse cell above
a horizontal coarse cell face.

\begin{table}[tbp]
\centering
\caption{Maximum velocity magnitude [m/s] in the box with a refined region after 10 time
steps, at the water level $z_w = 0.49$\,m. Columns as in Tab.~\ref{tab:nonOrthSkew}.
\figdir{tab_nonOrthRefined}.}
\label{tab:nonOrthRefined}
\small
\begin{tabular}{@{}lcccc@{}}
\toprule
 & \multicolumn{3}{c}{corrected} & uncorrected \\
 & $n_c = 0$ & $n_c = 2$ & hydrostatic & \\
\midrule
interFoam, basic                    & 1.0                & 1.0                & 1.0                 & $7.0\times10^{-1}$ \\
interIsoFoam, basic                 & 4.5                & 4.5                & 4.5                 & 3.3 \\
interIsoGoam, constantHeight        & $2.4\times10^{-2}$ & $5.5\times10^{-8}$ & $2.4\times10^{-9}$  & $3.4\times10^{-10}$ \\
interIsoGoam, geometric, $n_L = 1$  & 1.5                & 1.5                & 1.5                 & $7.5\times10^{-10}$ \\
interIsoGoam, geometric, $n_L = 2$  & $2.4\times10^{-2}$ & $1.8\times10^{-7}$ & $1.2\times10^{-10}$ & $7.7\times10^{-10}$ \\
\bottomrule
\end{tabular}
\end{table}

\parhead{The basic model.} Although no face is tilted, the basic model gives velocities of
order 1\,m/s with both corrected and uncorrected schemes (Tab.~\ref{tab:nonOrthRefined}).
The density jump is also carried by the split faces at the refinement boundary, whose
centres lie at different heights.

\parhead{The interface height.} The model \texttt{constantHeight} behaves as on the skewed box, and with
corrected schemes it has only the start-up error. The hydrostatic start eliminates this error,
and two non-orthogonal corrector loops reduce it to $5.5\times10^{-8}$\,m/s. Five and ten
loops give the same value.\datafn{run/refinedBox/data/nonOrthCorrectors_z0.49.txt} With
uncorrected schemes the geometric gravity model with one layer also keeps the fluid at rest,
but with corrected schemes it gives 1.5\,m/s, independently of the initialisation and the
number of non-orthogonal corrector loops (Tab.~\ref{tab:nonOrthRefined}).
The second condition for the discrete rest state, $gh_f = \grav\cdot\x_I$ on every face with
$\nabla_f\rho\neq 0$ (Eqn.~\ref{eq:nonOrthBalance}), is therefore violated on this mesh. On some
of these faces, one cell carries the inaccurate cell-centre height (Eqn.~\ref{eq:geometric}) instead of the
interface height.

\begin{figure}[tbp]
\centering
\begin{minipage}[t]{0.495\linewidth}
  \centering
  \includegraphics[width=\linewidth]{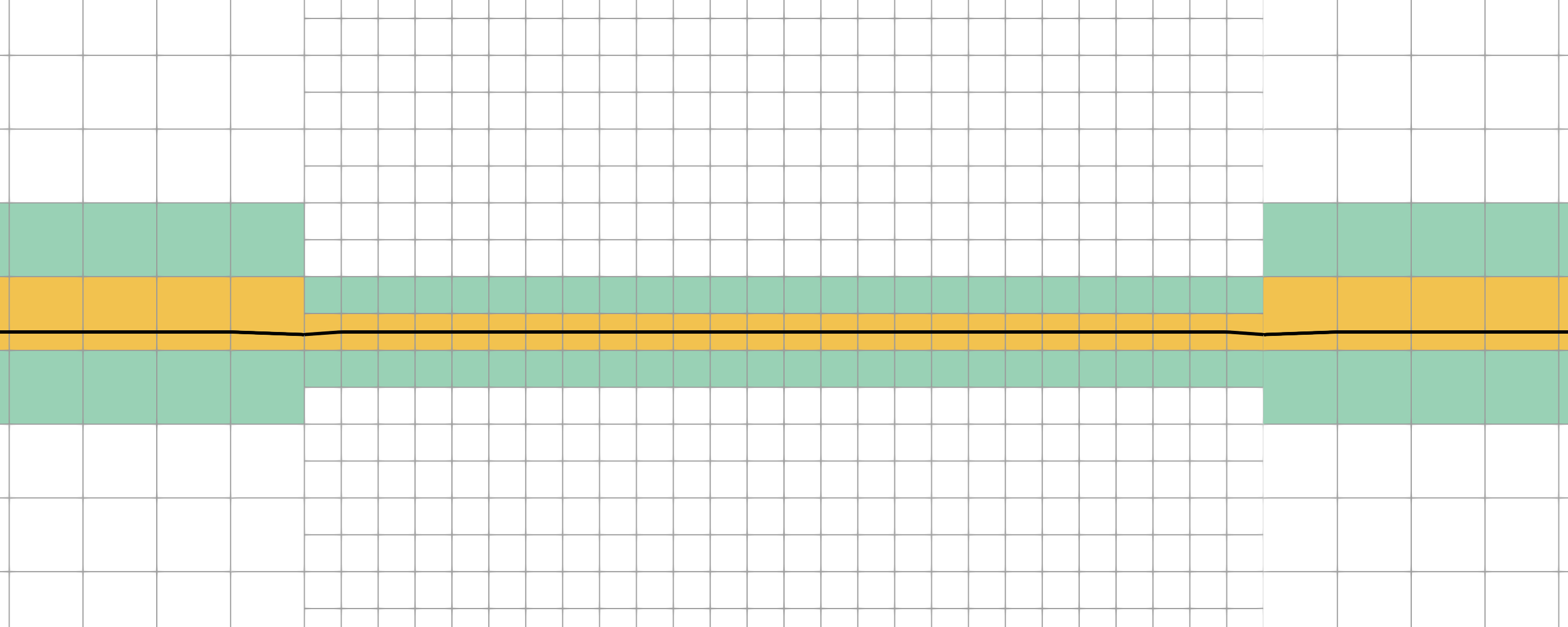}\\
  \small (a) uncorrected
\end{minipage}\hfill
\begin{minipage}[t]{0.495\linewidth}
  \centering
  \includegraphics[width=\linewidth]{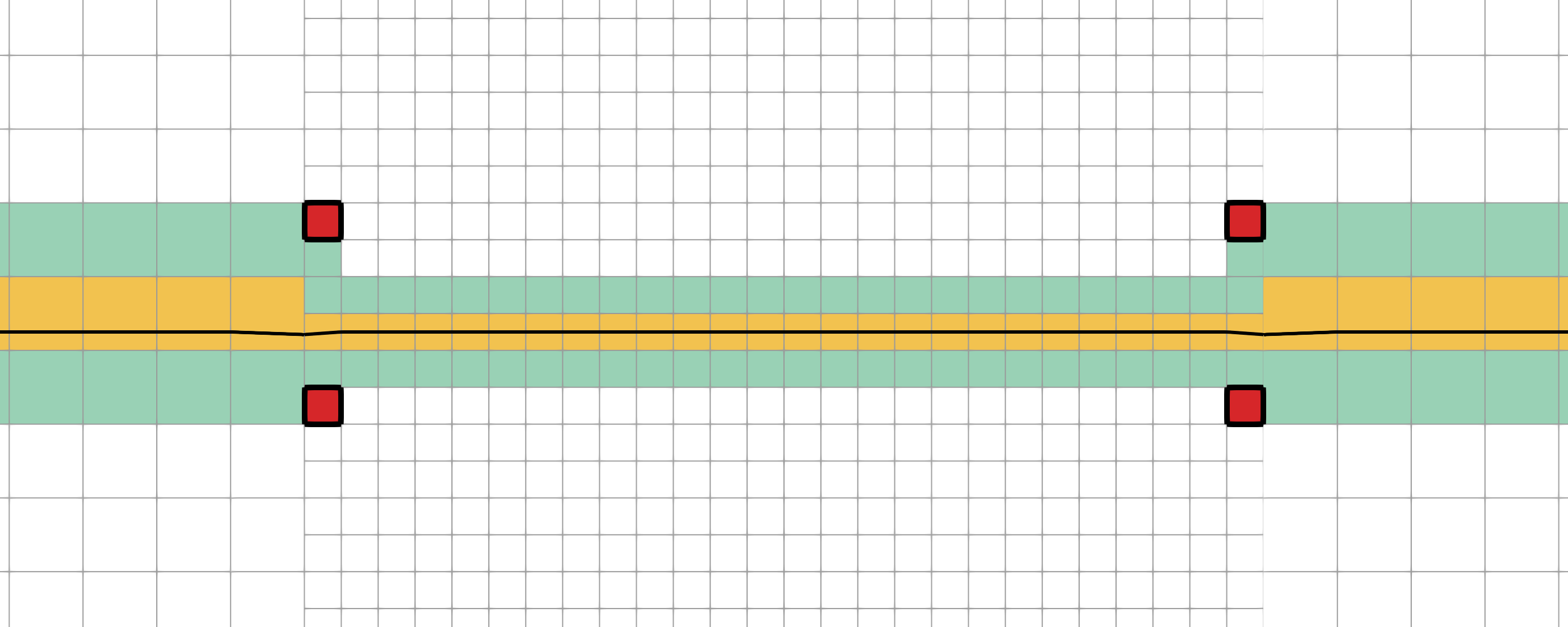}\\
  \small (b) corrected
\end{minipage}
\caption{Box with a refined region, $z_w = 0.49$\,m: the cells next to faces with a non-zero
density gradient, $|\nabla_f\rho|\,|\mathbf{d}_f| > 10^{-8}|\rho_1-\rho_2|$, in the initial
field, with (a) uncorrected and (b) corrected schemes. The colours give the layer, with interface cells
in amber, layer 1 in green and layer 2 in red with an outline. The black line is the contour $\alpha = 0.5$.
\figdir{fig_reach}.}
\label{fig:reach}
\end{figure}

\begin{figure}[tbp]
\centering
\includegraphics[width=0.9\linewidth]{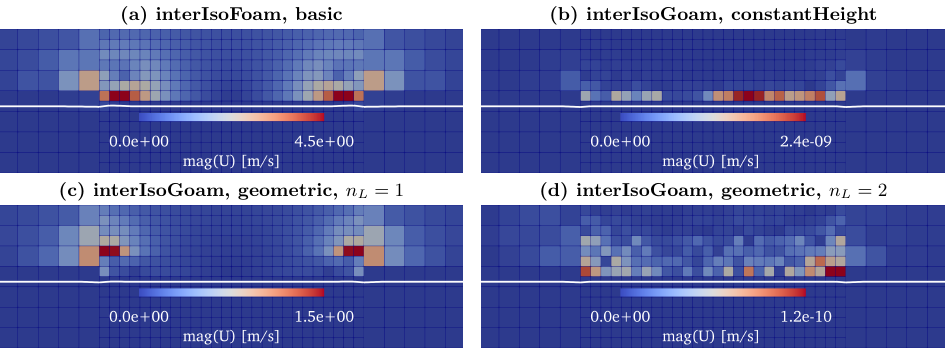}
\caption{Velocity magnitude in the box with a refined region for $z_w = 0.49$\,m after 10
time steps, from a hydrostatic initial $p_{rgh}$ with corrected schemes, in the strip
$0.36 < z < 0.64$\,m around the interface. Each panel has its own colour scale, from zero to
its maximum value, which is given in the panel. With one layer (c) the largest velocity occurs at the layer-2 cells of
Fig.~\ref{fig:reach}b. With two layers (d) the fluid stays at rest.
\figdir{fig_refinedBox}.}
\label{fig:refinedBox}

\end{figure}

\parhead{The reach of the corrected density gradient.} Without the non-orthogonal correction term,
$\nabla_f\rho\neq 0$ only on faces of interface cells, whose other cell is an interface cell or lies in layer 1.
With the correction, $\nabla_f\rho\neq 0$ also where the interpolated cell gradient has a
component along $\mathbf{k}_f$ in Eqn.~\ref{eq:corrected}. The cell gradient is non-zero in the interface cells and
their face neighbours, which belong to layer 1, so a face between a layer-1 cell and a layer-2
cell can carry a density gradient. In a cell with parallel opposite faces this contribution
vanishes, which is why the skewed box was not affected.\datafn{run/skewedBox/data/snGradRhoReach.txt}
Figure~\ref{fig:reach} shows the
cells adjacent to faces with a non-zero density gradient on the refined mesh. With uncorrected
schemes all of them are interface or layer-1 cells, while with corrected schemes four layer-2 cells at
the refinement boundary are added. With $n_L = 2$ these cells receive the interface height, and the
geometric gravity model behaves like \texttt{constantHeight} (Fig.~\ref{fig:refinedBox}d,
Tab.~\ref{tab:nonOrthRefined}).

\parhead{Summary.} A horizontal interface stays at rest on the skewed and the locally
refined mesh if the gravity term
uses the interface height in every cell that the density gradient reaches ($n_L = 2$), the
same face gradient is used for $p_{rgh}$ and $\rho$, and, with corrected schemes, $p_{rgh}$ is
initialised hydrostatically. Non-orthogonal corrector loops reduce the start-up error but do
not eliminate it.

\section{A standing wave across a refinement boundary}
\label{sec:standingWave}

Sections~\ref{sec:tiltedBox} and~\ref{sec:nonOrth} concerned a horizontal interface at rest,
for which an exact discrete balance exists. At a sloping interface the remaining gravity term
is not a gradient (Section~\ref{sec:theory}). The pressure cannot balance it, and the
unbalanced part is the restoring force of the wave. The loop sums of the discrete force (Eqn.~\ref{eq:loopSum}) should
then reproduce the circulation of the exact gravity term rather than vanish, so there is no
exact discrete balance to check. This section
measures instead how much of the improvement of Sections~\ref{sec:tiltedBox}
and~\ref{sec:nonOrth} remains for the moving interface of a standing wave.

\begin{table}[tbp]
\centering
\caption{Standing wave: maximum velocity magnitude after the first time step, its peak over
the first 20 steps and over the first two periods, and its time mean over the two periods.
Band: number of cells with $0 < \alpha < 1$ after one period. Linear theory gives 0.0154\,m/s after one
step and a peak of 0.290\,m/s. \figdir{tab_standingWave}.}
\label{tab:standingWave}
\small
\begin{tabular}{@{}llccccc@{}}
\toprule
 & & \multicolumn{2}{c}{$\max|\mathbf{U}|$ [m/s]} &
     \multicolumn{2}{c}{two periods [m/s]} & band \\
\cmidrule(lr){3-4}\cmidrule(lr){5-6}
Mesh & Solver & 1 step & peak, 20 steps & peak & mean & [cells] \\
\midrule
uniform & interFoam     & $0.943$ & $0.943$ & $0.94$ & $0.56$ & $135$ \\
        & interIsoFoam  & $0.943$ & $1.575$ & $1.97$ & $1.12$ & $63$ \\
        & interIsoGoam  & $0.199$ & $0.411$ & $0.71$ & $0.36$ & $62$ \\
\midrule
refined & interFoam     & $2.519$ & $2.519$ & $2.52$ & $0.84$ & $436$ \\
        & interIsoFoam  & $2.519$ & $4.048$ & $4.57$ & $2.41$ & $103$ \\
        & interIsoGoam  & $0.323$ & $1.645$ & $1.91$ & $0.84$ & $93$ \\
\bottomrule
\end{tabular}
\end{table}

\begin{figure}[tbp]
\centering
\includegraphics[width=0.7\linewidth]{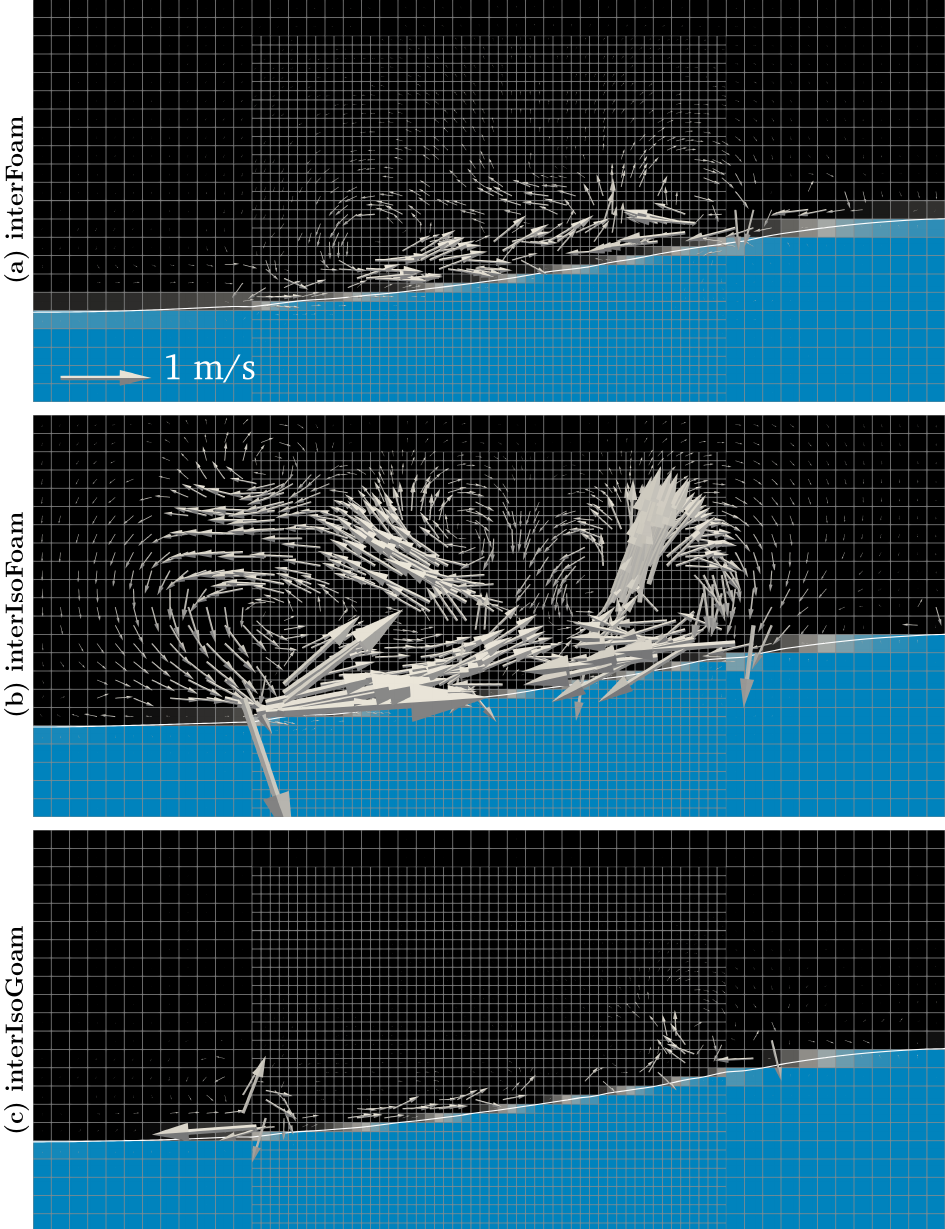}
\caption{Standing wave on the refined mesh after one period, when the velocity is supposed
to be approximately zero everywhere. Cells are coloured by volume fraction (air black,
water blue), with the contour $\alpha = 0.5$ in white. Arrows show the cell velocities on a common
scale, with a reference arrow of 1\,m/s in (a). \figdir{fig_standingWaveSnapshots}.}
\label{fig:standingWaveSnapshots}
\end{figure}

\parhead{Setup.} A standing wave with wavelength 2\,m (wave number $k = \pi$\,m$^{-1}$) and amplitude
$a = 0.05$\,m is released from
rest at maximum elevation in a closed box of $1\times1$\,m with water depth 0.5\,m. Linear
wave theory, which is exact only in the limit of small amplitude (here $ka = 0.16$), gives the
maximum velocity magnitude $u_{max}\sin(\omega t)$ with angular frequency $\omega = 5.317$\,s$^{-1}$ and
$u_{max} = 0.290$\,m/s.
After the first time step of 0.01\,s the linear-theory velocity is
0.0154\,m/s. The interface has not yet moved, so this step measures the gravity term. Two
meshes are used, a uniform $50\times50$ mesh, and the same mesh with the square
$0.24 \le x, z \le 0.76$\,m refined by a factor of two, so that the interface crosses the
refinement boundary at $x = 0.24$ and $0.76$\,m. All runs use the corrected schemes with two non-orthogonal corrector loops and $n_L = 2$,
and the volume fractions
are initialised with \texttt{setAlphaField}. Because of the refined mesh, the interface is
reconstructed with \texttt{plicRDF}, which is accurate on general meshes
\cite{Scheufler2019}.

\parhead{Results.} In the first time step \texttt{interFoam} and \texttt{interIsoFoam} give the same velocity,
as in the tilted box, and \texttt{interIsoGoam} reduces it by a factor of 4.7 on the uniform mesh
and 7.8 on the refined mesh (Tab.~\ref{tab:standingWave}). Its first-step velocity on the
uniform mesh is nevertheless about 13 times the value of 0.0154\,m/s expected from linear
theory. The refinement increases the first-step velocity of \texttt{interFoam} and
\texttt{interIsoFoam} by a factor of 2.7 and that of \texttt{interIsoGoam} by a factor of 1.6.
The geometric gravity model therefore reduces the effect of the refinement boundary but does not remove
it. Over two periods, \texttt{interIsoGoam} reduces the peak and the mean velocity of
\texttt{interIsoFoam}, which differs from it only in the gravity term, by factors of 2.4 to 3.1
on both meshes. Its mean velocity is lower than that of \texttt{interFoam} on the uniform mesh
(0.36 against 0.56\,m/s) and equal to it on the refined mesh (0.84\,m/s), although
the interface band of \texttt{interFoam} is two to five times wider. The geometric gravity model
thus reaches or improves on the velocity level of the algebraic method while keeping the
interface sharp. After one period the remaining spurious
velocities of \texttt{interIsoGoam} are confined to the air near the two points where the interface
crosses the refinement boundary (Fig.~\ref{fig:standingWaveSnapshots}).

\parhead{Air and water.} In every run the largest velocity is in the air ($\alpha < 0.01$). In
the water ($\alpha > 0.99$), \texttt{interIsoGoam} gives 0.0151 and 0.0153\,m/s after the first
step on the uniform and the refined mesh, close to the linear value of 0.0154\,m/s, whereas
\texttt{interFoam} and \texttt{interIsoFoam} give 0.0180\,m/s on the refined mesh. Over two
periods the largest water velocity of \texttt{interIsoGoam} is 0.30 and 0.32\,m/s on the two
meshes, against 0.41 and 0.44\,m/s for \texttt{interFoam} and \texttt{interIsoFoam} on the
refined mesh and 0.290\,m/s from linear theory. The remaining error of the geometric gravity
model at a sloping interface is therefore a spurious velocity in the air, while the water
moves close to the linear solution.\datafn{papers/part1/tab_standingWave/tab_standingWave.txt}

\section{A progressive wave in a periodic domain}
\label{sec:wave}

\begin{figure}[tbp]
\centering
\includegraphics[width=\linewidth]{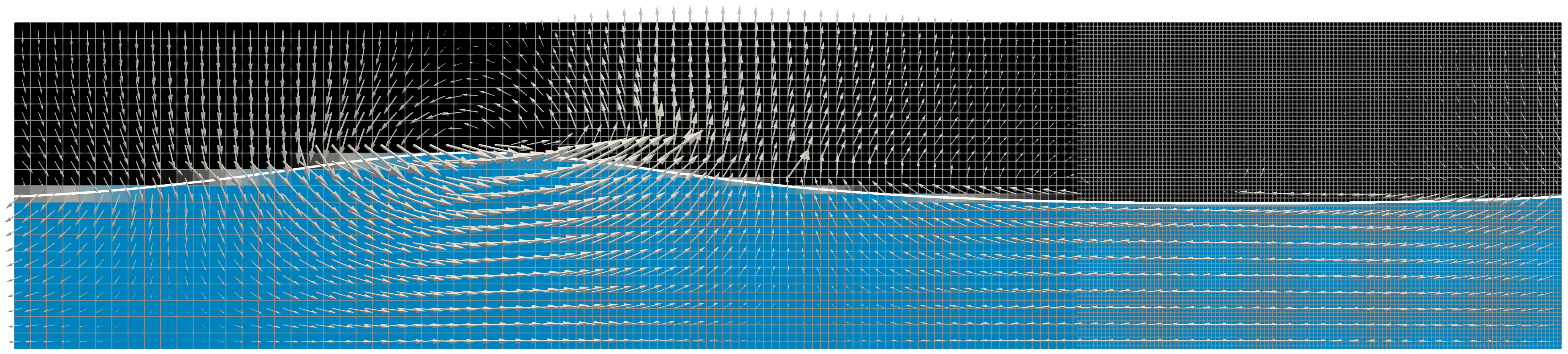}
\caption{The periodic wave tank, one wavelength (3.79\,m) long, at $t = 0.2$\,s. The
left, middle and right thirds show the $95\times20$, $190\times40$ and $380\times80$
meshes. Cells are coloured by volume fraction (air black, water blue). The white line is the
free surface on the $380\times80$ mesh, and the arrows show the velocity on the $95\times20$ mesh (\texttt{interIsoGoam}, maximum Courant number 0.5). \figdir{fig_waveSetup}.}
\label{fig:waveSetup}

\end{figure}

\parhead{Setup.} A nonlinear progressive wave with height $H = 0.125$\,m and period
$T = 2$\,s in water of depth 0.4\,m (wavelength $L = 3.79$\,m, steepness $H/L = 0.033$) is
propagated in a periodic domain of one wavelength that extends 0.4\,m above the still water
level (Figs.~\ref{fig:cases}e and~\ref{fig:waveSetup}). The wave, the domain and the
$380\times80$ mesh follow Larsen et al.\ \cite{Larsen2019}, who chose the wave to match the
incoming wave of the spilling breaker experiment of Ting and Kirby \cite{TingKirby1994}. The
exact solution and the initial fields are computed with stream function theory
\cite{Rienecker1981,Fenton1988} using the streamFunctionWave utility
\selfcite{streamFunctionWave}. We call this solution the exact input wave. Stream function theory
describes only the water, so the air is initialised at rest. Four meshes of square cells are used, from $95\times20$ to
$760\times160$ cells, that is 3 to 25 cells per wave height. The densities are 1000
and 1\,kg/m$^3$.

All runs use Euler time stepping, one outer and three inner pressure correctors without
momentum predictor, and limited linear interpolation for momentum convection. All three solvers include the \texttt{ddtCorr} term of \OF, as
\texttt{interFoam} and \texttt{interIsoFoam} always do (Appendix~\ref{app:ddtCorr}). As a moving interface has no exact balance to preserve
and the meshes are uniform, the \texttt{interIsoFoam} and
\texttt{interIsoGoam} runs reconstruct the interface with \texttt{isoAlpha}
\cite{Roenby2016}, the fastest reconstruction of Scheufler and Roenby \cite{Scheufler2019}.
\texttt{interIsoGoam} uses the geometric gravity model (Eqn.~\ref{eq:geometric}) with the
default $n_L = 2$. Two sets of runs, each over 15 periods
(30\,s), are made. In the first, all solvers use the same fixed time steps, from a largest time
step $\Delta t_0$, halved three times, resulting in maximum Courant numbers in the range from roughly 0.5
to 0.04. In the
second set of runs, the time step is set by a Courant limit of 0.5, 0.2, 0.1 or 0.05.

Two quantities measure the accuracy. The surface elevation is recorded at a fixed gauge
with the function object \texttt{surfacePosition} \selfcite{surfacePosition}, and each wave height
is the crest height minus the height of the following trough. To reduce the scatter from wave
to wave, a straight line is fitted through the last five wave heights. The wave height error
$\Delta H$ is the value of this line at the last wave, relative to the exact wave height. The maximum velocity magnitude in the domain is recorded
every time step. In the exact solution it is the crest velocity $u_{crest} = 0.445$\,m/s.
Its median over $t > 2$\,s, relative to $u_{crest}$, measures the spurious velocities.

\begin{figure}[tbp]
\centering
\includegraphics[width=0.85\linewidth]{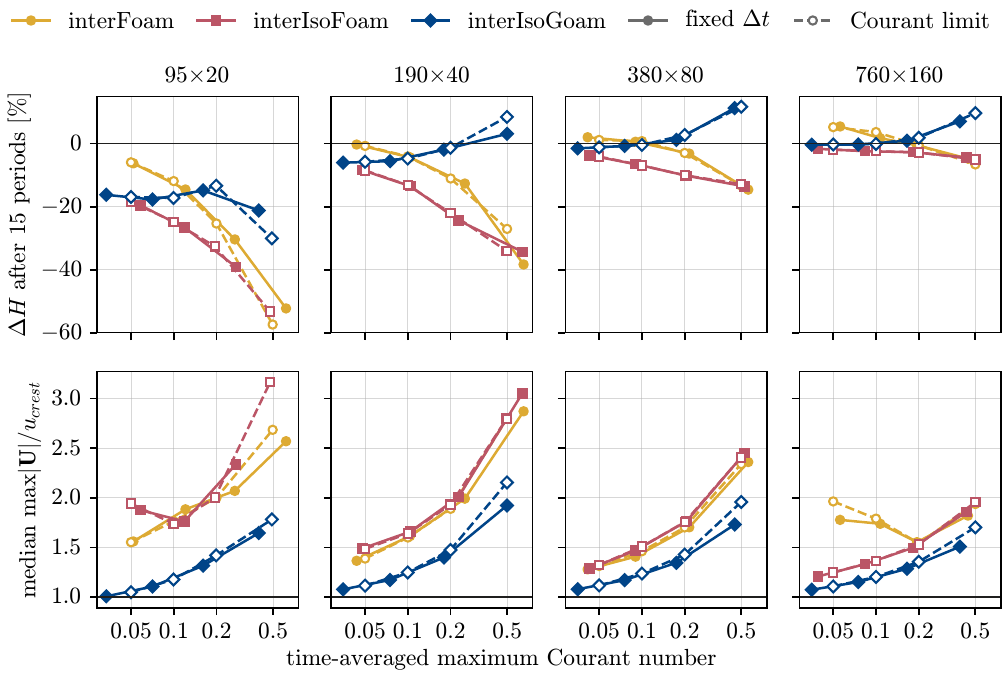}
\caption{Progressive wave, all runs, against the time-averaged maximum Courant number, with one
column per mesh. Top: wave height error after 15 periods. Bottom: median over time of the
maximum velocity magnitude in the domain, relative to the crest velocity (1 for the exact
solution). Solid lines with filled markers are runs with a fixed time step, from $\Delta t_0$ to
$\Delta t_0/8$ from right to left, and dashed lines with open markers are runs under a Courant limit. The diverged run of
\texttt{interIsoFoam} ($95\times20$, $\Delta t_0$) is omitted.
\figdir{fig_waveResults}.}
\label{fig:waveResults}
\end{figure}

\begin{figure}[tbp]
\centering
\includegraphics[width=0.55\linewidth]{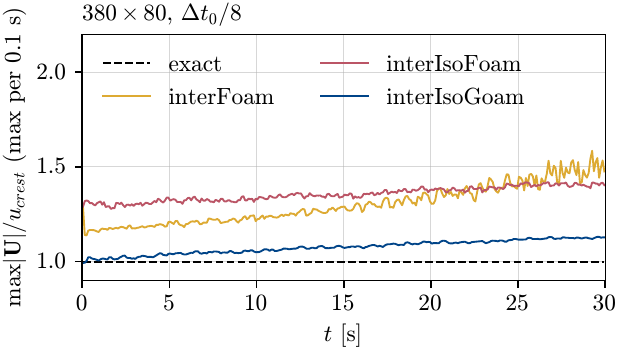}
\caption{Progressive wave on the $380\times80$ mesh at $\Delta t_0/8$: maximum velocity
magnitude in the domain relative to the crest velocity, largest value in each 0.1\,s
interval. The medians of the three solvers are in the same order as in most runs
(Fig.~\ref{fig:waveResults}).
\figdir{fig_spurious}.}
\label{fig:spurious}

\end{figure}

\begin{figure}[tbp]
\centering
\includegraphics[width=0.85\linewidth]{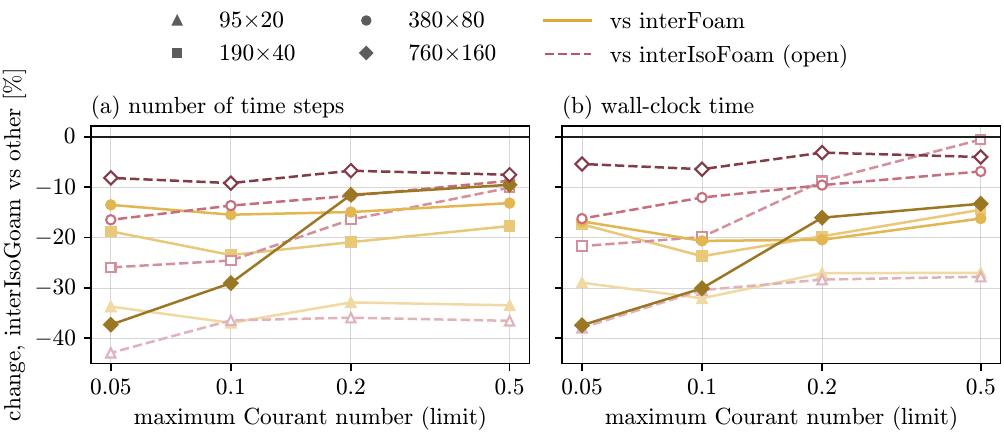}
\caption{Progressive wave under a Courant limit: change in (a) the number of time steps and
(b) the wall-clock time of \texttt{interIsoGoam} relative to \texttt{interFoam} (gold, solid,
filled markers) and \texttt{interIsoFoam} (rose, dashed, open markers). Marker shapes
distinguish the meshes, and darker shades are finer meshes. Negative values mean that
\texttt{interIsoGoam} needs less. The wall-clock time is the number of steps times the time per step of
separate timing runs, taken as the median over two repeats (three on $760\times160$).
\figdir{fig_waveCost}.}
\label{fig:waveCost}
\end{figure}

\parhead{Spurious velocities in the air.} \texttt{interIsoGoam} has the lowest maximum
velocity of the three solvers in every run, with fixed time steps and under a Courant limit
(Fig.~\ref{fig:waveResults}, bottom row). At the smallest
time step on the three finer meshes, its maximum velocity exceeds the crest velocity by at
most 8\,\%, against 21--49\,\% for \texttt{interIsoFoam} and 28--78\,\% for \texttt{interFoam}. Between \texttt{interFoam} and \texttt{interIsoFoam} the ordering varies: \texttt{interFoam} has
the lower maximum velocity of the two in most runs, and a clearly higher one only on the finest mesh at
the smallest time steps (Fig.~\ref{fig:waveResults},
bottom row). Figure~\ref{fig:spurious} shows for one run that the maximum velocity of all three
solvers grows slowly over the 15 periods, and that \texttt{interIsoGoam} stays lowest
throughout. Since the time step under a Courant limit is set by the largest
velocity in the domain, \texttt{interIsoGoam} needs 7 to 43\,\% fewer time steps than the other
solvers (Fig.~\ref{fig:waveCost}a). The largest savings
relative to \texttt{interFoam}, on the finest mesh at the two smallest Courant limits, occur in the runs
in which the crests of \texttt{interFoam} become irregular (see below). The wall-clock time falls by up
to 38\,\%, and in no run is \texttt{interIsoGoam} slower (Fig.~\ref{fig:waveCost}b). At the
maximum Courant numbers of 0.1 and below recommended in Section~\ref{sec:conclusion}, it needs
8 to 43\,\% fewer steps and 5 to 38\,\% less wall-clock time.\datafn{papers/part1/fig_waveCost/waveCost.txt} Relative
to \texttt{interIsoFoam} it falls less than the number of steps, because a time step of
\texttt{interIsoGoam} costs 3 to 13\,\% more, most on the coarsest mesh. The wall-clock time
is the number of steps of each run times the time per step, measured in separate, shorter runs
of the same set-ups. In these runs the solvers compared ran at the same time on an otherwise
idle machine, each process bound to its own core and its memory to the NUMA node of that core,
and repeated runs give changes within
3 percentage points of each other.\datafn{run/progressiveWave/timing_partI} The only run that diverged is
\texttt{interIsoFoam} on the coarsest mesh at the largest fixed time step, while the corresponding run under a Courant
limit of 0.5 completed.

\parhead{The wave height.} At maximum Courant numbers of about 0.1 and below the wave height error of
\texttt{interIsoGoam} depends little on the time step and decreases steadily with mesh
refinement, to $-0.3$\,\% on the finest mesh at the smallest time step
(Fig.~\ref{fig:waveResults}, top row). The error of \texttt{interIsoFoam} decreases with both
the time step and the mesh, but more slowly, to $-1.7$\,\%. That of \texttt{interFoam} changes
sign on the two finest meshes as the Courant number is reduced and reaches $+5.5$\,\%. At the
largest time step (maximum Courant number about 0.5) on the $380\times80$ mesh,
\texttt{interFoam} and \texttt{interIsoFoam} damp the wave by 14 to 15\,\% over 15 periods,
whereas \texttt{interIsoGoam} increases its height by 11\,\%. Runs with a Courant limit give nearly the same errors
as runs with a fixed time step at the same average Courant number, except at the largest
Courant numbers on the two coarsest meshes.

\begin{figure}[tbp]
\centering
\includegraphics[width=0.8\linewidth]{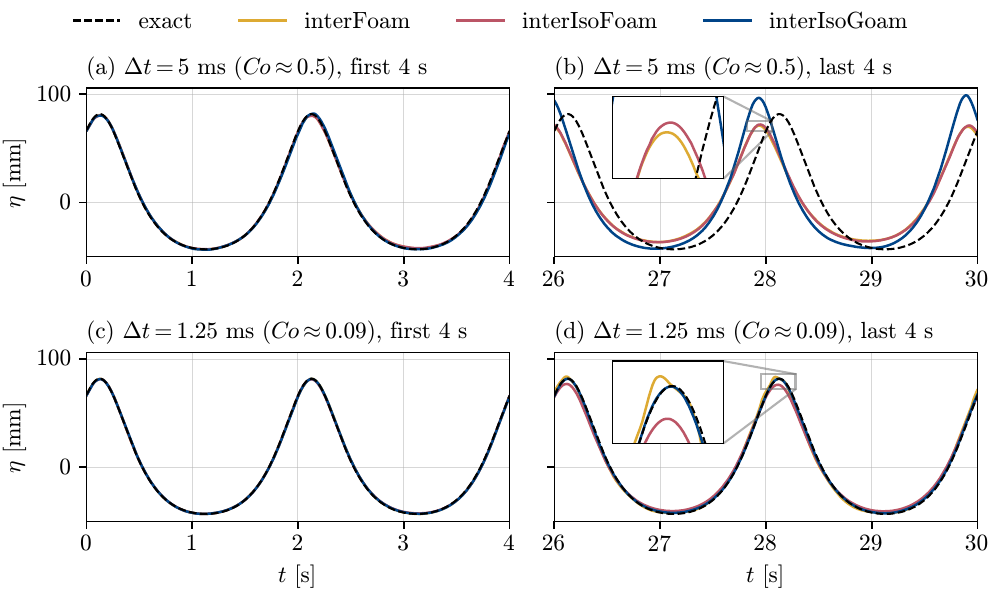}
\caption{Surface elevation at the gauge on the $380\times80$ mesh during the first two and
the last two periods, at the largest time step $\Delta t_0$ (a, b) and at $\Delta t_0/4$
(c, d). The dashed line is the exact solution. The insets in (b) and (d) enlarge the grey box
around the crest near $t = 28$\,s.
\figdir{fig_eta}.}
\label{fig:eta}
\end{figure}

Figure~\ref{fig:eta} shows the surface elevation on the $380\times80$ mesh at $\Delta t_0$
and $\Delta t_0/4$. During the first periods all three solvers follow the exact solution at
both time steps. At the largest time step, after 15 periods, \texttt{interFoam} and
\texttt{interIsoFoam} have damped the wave and \texttt{interIsoGoam} has amplified it, mostly at
the crests. The crests of all three solvers arrive early by about the same amount, so the error
in the wave speed is not caused by the gravity term. At $\Delta t_0/4$ this error has almost
disappeared, which shows that it is set by the time step. At $\Delta t_0/4$ (maximum
Courant number about 0.09) all three solvers stay close to the exact solution,
although \texttt{interIsoFoam} has slightly lower crests. The wave height error of \texttt{interIsoGoam}
is then $-0.7$\,\%.

\begin{figure}[tb]
\centering
\includegraphics[width=0.8\linewidth]{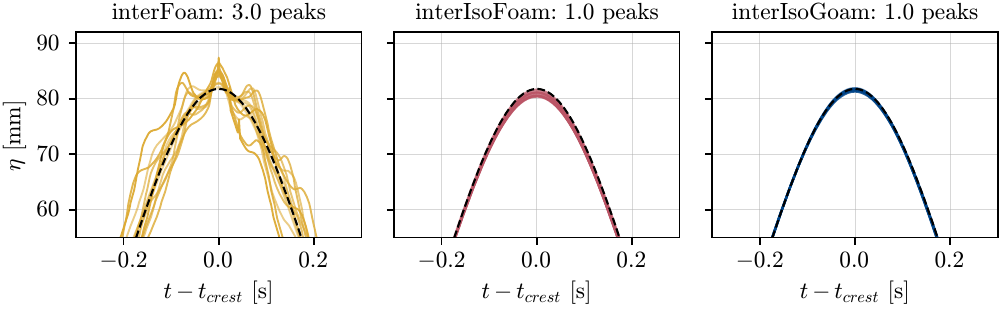}
\caption{Surface elevation at the gauge around every crest after the first, against the time
$t - t_{crest}$ from the highest elevation of that crest, on the $760\times160$ mesh at
$\Delta t_0/8$. Later crests are darker, and the dashed line is the exact solution. The panel
titles give the mean number of peaks per crest.
\figdir{fig_crests}.}
\label{fig:crests}
\end{figure}

\parhead{The crest shape.} On the finest mesh, at the two smallest fixed time steps and the two smallest Courant
limits, and only there, the crests of \texttt{interFoam} become irregular. Each crest then carries two
to three peaks, and its shape changes from wave to wave (Fig.~\ref{fig:crests}). In the run
shown, the wave height of \texttt{interFoam} grows by 5.5\,\% (Fig.~\ref{fig:waveResults}). The crests of \texttt{interIsoFoam} and \texttt{interIsoGoam} have a single peak in all
runs. In the run of Fig.~\ref{fig:crests}, the crests of \texttt{interIsoGoam} coincide with the
exact crest and those of \texttt{interIsoFoam} lie slightly below it.

\begin{figure}[tb]
\centering
\includegraphics[width=0.85\linewidth]{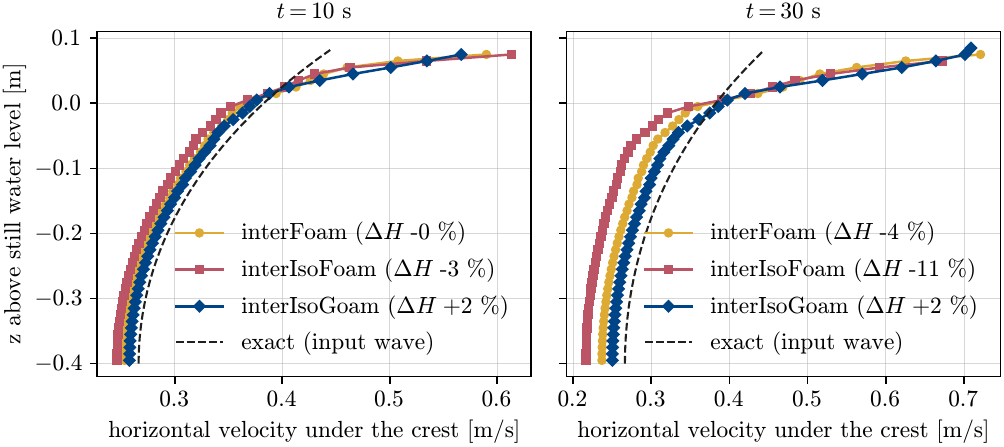}
\caption{Progressive wave on the $380\times80$ mesh, maximum Courant number 0.2: horizontal
velocity under the crest at $t = 10$ and 30\,s, after about 5 and 15 periods. The dashed
line is the exact input wave, and $\Delta H$ is the change of the wave height at that time.
\figdir{fig_crestProfiles}.}
\label{fig:crestProfiles}
\end{figure}

\begin{figure}[tbp]
\centering
\includegraphics[width=0.8\linewidth]{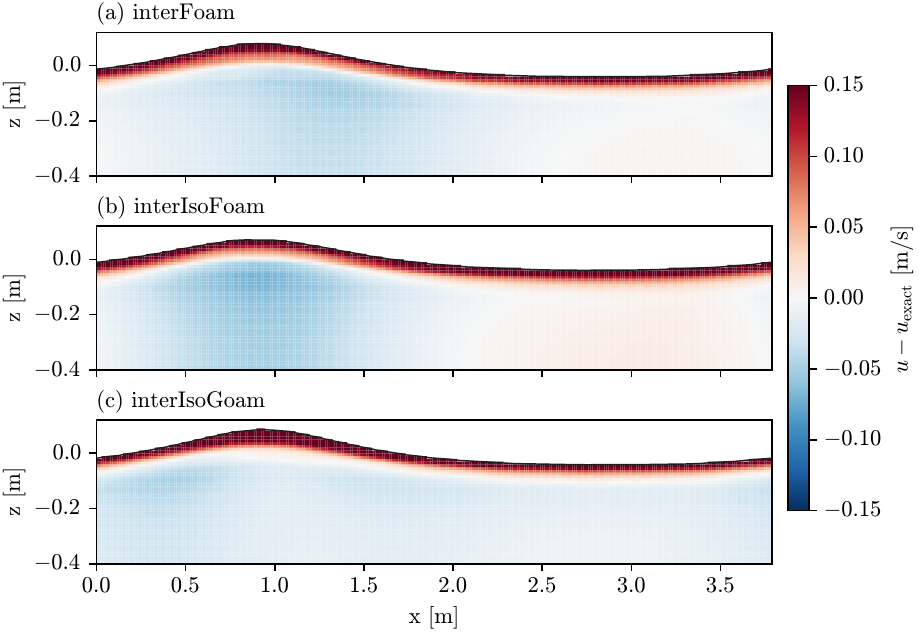}
\caption{Progressive wave on the $380\times80$ mesh, maximum Courant number 0.2: error of
the horizontal velocity in the water after 15 periods, relative to the exact input wave
shifted to the crest position of each run (air not shown). The black line is the contour
$\alpha = 0.5$. \figdir{fig_waveField}.}
\label{fig:waveField}

\end{figure}

\parhead{The crest velocity.} All three solvers overpredict the
horizontal velocity near the surface under the crest and underpredict it further down
(Fig.~\ref{fig:crestProfiles}). The velocity error field shows a band of excess velocity
along the whole surface in all three solvers (Fig.~\ref{fig:waveField}). The overprediction
in the top full water cell grows with time, from 18--29\,\% after 5 periods to 45--60\,\% after 15 periods, and it appears with both the basic and the geometric gravity
model, slightly larger with the latter.\datafn{papers/part1/tab_crest/tab_crest.txt} It is therefore not caused by
the gravity term. Part~II \selfcite{RoenbyPart2} shows that it originates in the
momentum convection term.

\section{Conclusions}
\label{sec:conclusion}

The \OF solvers \texttt{interFoam} and \texttt{interIsoFoam} for incompressible two-phase flows
absorb the gradient part of the gravity force,
$\nabla(\rho\,\grav\cdot\x)$, into the auxiliary pressure $p_{rgh}$. The remaining gravity term
acts only where the density jumps, as a force on the cell faces proportional to the
height at which the density jumps. Both solvers evaluate this height at the
face centre. The resulting force discretisation error is of the
order of the density jump times the gravitational acceleration, independent of the cell
size, and on meshes not aligned with gravity the pressure cannot balance it. Evaluating the height at the centroids of the interface segments reconstructed
by a geometric VoF method makes the discrete force an exact discrete gradient for a horizontal
interface on any mesh.

We tested the model in two dimensions on cases of increasing complexity: a horizontal interface
at rest on tilted, skewed and locally refined meshes, where the exact solution is known, and
then a standing and a progressive wave. The main findings are the following.

\begin{itemize}
\item For a horizontal interface at rest in a tilted box, the basic gravity term causes a spurious
acceleration of several times gravity in the first time step, independent of the time step
and not decreasing with mesh refinement. The geometric gravity model reduces the velocities
from metres per second to the order of $10^{-9}$\,m/s, close to the values with the exact interface height, and keeps this advantage
for density ratios from 10 to $10^7$. Its residual velocity increases in inverse proportion to
the density of the light fluid.
\item On non-orthogonal meshes the rest state remains an exact discrete solution if the
same face gradient scheme is used for $p_{rgh}$ and $\rho$ and the interface height reaches
every cell touched by the non-orthogonal correction of the density gradient. Near a mesh
refinement this requires a second layer of cells ($n_L = 2$). With corrected schemes and a
zero initial $p_{rgh}$, a start-up error remains, which a hydrostatic initial $p_{rgh}$
eliminates and non-orthogonal corrector loops reduce.
\item For moving interfaces the spurious velocities are reduced but not removed. For a
standing wave on a uniform mesh and across a refinement boundary, the geometric gravity model
reduces the error of the first time step five- to eightfold, and the spurious velocities over
two periods by a factor of 2.4 to 3.1, relative to \texttt{interIsoFoam}. Its first-step velocity
remains well above the value expected from linear theory.
\item For a progressive wave, the model gives the lowest maximum velocity of the three solvers
in every run and needs 7 to 43\,\% fewer time steps and up to 38\,\% less wall-clock time under a
Courant limit. At maximum Courant numbers of about 0.1 and below the wave height
error of the model decreases steadily with mesh refinement, to $-0.3$\,\% on the finest mesh.
The overprediction of the crest velocity appears in all three solvers and is not caused by the
gravity term.
\end{itemize}

\parhead{Recommendations.} The conditions for the discrete rest state translate directly into
recommendations for users of the geometric gravity model. The model requires a geometric VoF
method, since only such a method provides the interface position. The face-normal gradient of
the density (\texttt{snGradSchemes} in \texttt{fvSchemes}) and the Laplacian of $p_{rgh}$
(\texttt{laplacianSchemes}) must use the same non-orthogonal correction, for example
\texttt{corrected} in both. The interface height must reach every cell that enters the density
gradient, and the default of two layers ensures this on all meshes tested here. With corrected
schemes, $p_{rgh}$ should be initialised hydrostatically. Non-orthogonal corrector loops
reduce the start-up error but do not remove it. For wave propagation the maximum Courant
number should be about 0.1 or less (Section~\ref{sec:wave}). These recommendations reflect the
two-dimensional meshes and tests of this paper. The model and the solvers are likely to be developed further, so users
should check the code repository and later publications for updated settings.

The progressive wave also shows two further effects.
With the geometric gravity model, the wave height grows by up to 12\,\% over 15 periods at a
maximum Courant number of 0.5 on the three finer meshes, but not at 0.1 and below. The origin of this growth is not
identified. Switching off the \texttt{ddtCorr} term,
which \texttt{interIsoGoam} allows, reduces the damping of the wave (Appendix~\ref{app:ddtCorr}).

\parhead{Open questions.} For a moving, sloping interface the discrete force must reproduce
the circulation of the exact gravity term, which provides the restoring force of the wave.
How to achieve this is open, and residual spurious velocities remain in the
light fluid near the interface. Part~II \selfcite{RoenbyPart2} traces the overprediction of the crest velocity to the
momentum convection term, whereas the question why the wave height grows at large Courant numbers remains
open. The layers are defined through point neighbours and so apply to any cell
shape, but the tests of this paper are two-dimensional. The performance on three-dimensional
unstructured meshes, including the effect of the reconstruction on the interface height,
remains to be tested in future work.

%-----------------------------------------------------------------------
%      Start with acknowledgements here
%-----------------------------------------------------------------------

\ifdefined\review
\else
  \section*{Acknowledgements}

% Acknowledgements (only included after the review process, before publication).
The InterFlow project, financed by Independent Research Fund Denmark
through a DFF Sapere Aude Research Leader grant (grant number 9063-00018B), partially funded
J.R.\ and the PhD studies of K.Mi.\ and K.M{\o}. We thank Benjamin Bouscasse for
drawing our attention to the effect of the \texttt{ddtCorr} term on wave propagation.

  % The OpenFOAM Journal requires the contributions of all authors to be explicitly stated.
  % Please see http://img.mdpi.org/data/contributor-role-instruction.pdf (CRediT taxonomy)
  % for the term explanation.
  \authorcontributions{
  Conceptualisation, J.R., H.S., K.Mi.\ and K.M{\o}.;
  methodology, J.R.\ and H.S.;
  software, J.R.\ and H.S.;
  validation, J.R., K.M{\o}.\ and S.A.;
  formal analysis, J.R.\ and K.M{\o}.;
  investigation, J.R., H.S., K.M{\o}.\ and S.A.;
  visualisation, J.R.;
  writing---original draft preparation, J.R.;
  writing---review and editing, all authors;
  supervision, J.R.;
  project administration, J.R.;
  funding acquisition, J.R.
  All authors have read and agreed to the published version of the manuscript.
  }

  \dataavailability{
  The solver, the case set-ups and the post-processing scripts are available at
  \url{https://github.com/isoAdvector/fluxinator}. The repository will be archived at Zenodo
  on publication.
  }
\fi

%    Where the reviewers find the code: the supplementary file (review version only)
\ifdefined\review
  \dataavailability{
  The solver, the case set-ups and the post-processing scripts are in the supplementary file,
  whose \texttt{README\_REVIEW.md} explains how to build and run them. The paths in the captions
  (\emph{Files}) and footnotes (\emph{Data}) refer to it. The repository will be archived at Zenodo
  on publication.
  }
\fi

%    Use of generative AI (shown also in the review version)
\section*{Declaration of generative AI use}

\noindent The authors used Claude Code with Anthropic's Claude Opus models extensively
in all aspects of this work, including the development of the solver and utilities, the
setup, running and post-processing of the test cases, the analysis of the results, the
preparation of figures and tables, the search for and checking of references, and the
drafting and editing of the text. All text, code and cases produced with the tool have
been thoroughly reviewed by the authors, who take full responsibility for the content of
this article, including all results and claims.

%-----------------------------------------------------------------------
%      Start with appendices here
%-----------------------------------------------------------------------

\appendix
\section{The \texttt{ddtCorr} term}
\label{app:ddtCorr}

\parhead{Origin.} On a collocated mesh the predicted face flux is interpolated from cell
values. To suppress checkerboard pressure modes, Rhie and Chow \cite{RhieChow1983} added a pressure term to this interpolation, which
couples the face flux to the pressure in the two cells of the face. In a transient solver, the interpolated flux is dominated at small time steps
by the old cell velocity, which enters through the time derivative. The face flux is then
reset to the interpolated old cell velocity in every time step, steady solutions depend on
the time step, and the pressure coupling weakens as $\Delta t \to 0$ \cite{Bartholomew2018}.
Choi \cite{Choi1999} removed this dependence by letting the face flux keep its own old value.
The \texttt{ddtCorr} term of \OF serves the same purpose. It adds to the predicted face flux the term
\begin{equation}
  c\,\frac{(\rho\,r_A)_f}{\Delta t}\bigl(\phi^n - \mathcal{F}(\U^n)\bigr),
  \qquad
  c = 1 - \min\left(\frac{|\phi^n - \mathcal{F}(\U^n)|}{|\phi^n|},\, 1\right),
  \label{eq:ddtCorr}
\end{equation}
where $\phi^n$ is the face flux of the previous time step, $\mathcal{F}(\U^n)$ the flux of
the interpolated cell velocity of that time step, $(\rho\,r_A)_f$ the interpolated product of
the density and the reciprocal $r_A$ of the diagonal of the momentum matrix, and $c$
the coefficient for Euler time stepping. The face flux then keeps part of its own old value, instead of
being rebuilt from the cell velocity in every time step. Vuorinen et al.\ \cite{Vuorinen2014}
documented the term, with this coefficient, in the PISO solver of \OF, but it is not part of
the PISO algorithm of Issa \cite{Issa1986}. Wave simulations have also been run without the term at
\'Ecole Centrale de Nantes. The numerical model of Descamps \cite{Descamps2022} does not
include it, and Tierno \cite{Tierno2026} reports its effect on wave propagation.

\parhead{Effect in wave simulations.} In wave simulations the motion is
oscillatory, the time step is set by a Courant limit, and the wave must
keep its energy over many periods. Since $c < 1$, only part of the difference between the old face flux and the
interpolated old cell velocity is kept, so in every time step the face flux is drawn towards
the interpolated cell velocity. This relaxation damps the flow and removes kinetic energy. Even in the consistent form of Choi, the interpolation removes kinetic energy at a
rate independent of the time step \cite{Bartholomew2018}. Vuorinen et al.\
\cite{Vuorinen2014} tested the term in an inviscid Taylor--Green vortex, for which the exact
kinetic energy is constant. With the term, the single-phase PISO solver of \OF lost 20\,\% of
the kinetic energy within two vortex time scales at a maximum Courant number of 0.5 on a
$32^2$ mesh and 2\,\% on a $128^2$ mesh, and less at smaller time steps. Without
it, the solver was comparable to explicit Runge--Kutta projection methods, which they found
nearly free of dissipation. Switching the term off may in turn weaken the suppression of
pressure oscillations at very small time steps \cite{Bartholomew2018}.

\begin{figure}[tbp]
\centering
\includegraphics[width=0.95\linewidth]{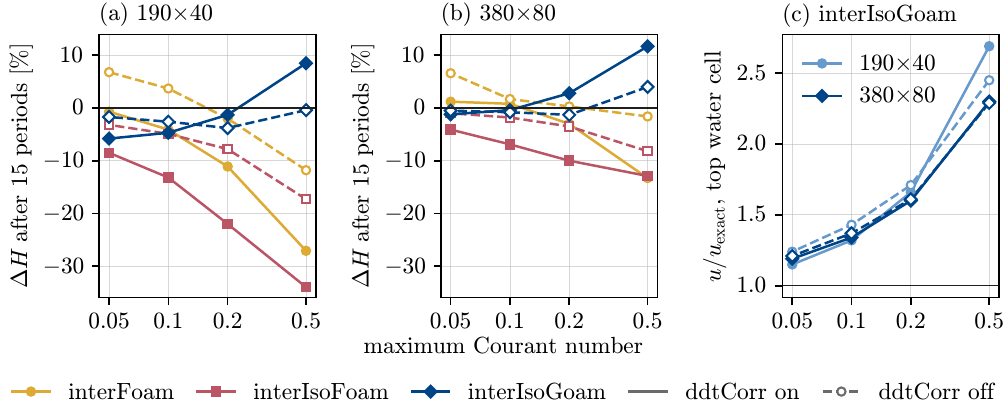}
\caption{Progressive wave under a Courant limit, \texttt{ddtCorr} term on (solid) and off
(dashed). (a), (b) Wave height error after 15 periods on two meshes. (c) \texttt{interIsoGoam}: horizontal velocity
in the top full water cell under the crest relative to the exact input wave, averaged over
28--30\,s, on the $190\times40$ (light) and $380\times80$ (dark) meshes.
\figdir{fig_ddtCorr}.}
\label{fig:ddtCorr}

\end{figure}

\parhead{The progressive wave.} Figure~\ref{fig:ddtCorr} repeats the runs of
Section~\ref{sec:wave} under a Courant limit on the two intermediate meshes with the
\texttt{ddtCorr} term switched off and everything else unchanged. With the term off, the errors of
\texttt{interIsoFoam} and \texttt{interIsoGoam} decrease steadily as the time step is reduced, and on the
$380\times80$ mesh they reach about 1\,\% or less. Switching the term off
clearly reduces the damping of \texttt{interIsoFoam}, and it also improves \texttt{interIsoGoam},
although less. The term thus
adds a damping that decreases slowly, if at all, as the time step is reduced, like the
dissipation in the Taylor--Green vortex \cite{Vuorinen2014}. The solver \texttt{interFoam} behaves differently. With the term off, its wave height
grows at small time steps, by about 7\,\% at the smallest. Part of its accuracy at small time
steps with the term on therefore comes from the damping of the term. The term has little
effect on the crest velocity (Fig.~\ref{fig:ddtCorr}c).

% The figure before the references
\FloatBarrier

%    Bibliographies can be prepared with BibTeX using IEEEtran style,
\bibliographystyle{IEEEtran}%do not change

\bibliography{Bibliography}

\end{document}